\documentclass[runningheads]{llncs}
\usepackage[T1]{fontenc}
\usepackage{amsmath,amssymb,amsfonts,yfonts}
\usepackage{graphicx}
\usepackage[table]{xcolor}
\usepackage{hyperref}
\usepackage{subcaption}
\usepackage{multirow}
\usepackage{enumitem}
\usepackage{todonotes}

\def\ME {\mathcal{E}}
\def\MA {\mathcal{A}}
\def\MV {\mathcal{V}}
\def\MI {\mathcal{I}}

\def\MR {\mathcal{R}}
\def\MS {\mathcal{S}}
\def\MM {\mathcal{M}}

\begin{document}
\title{Online Discovery of Simulation Models for Evolving Business Processes}
%
% \titlerunning{Online Discovery of Simulation Models}
\author{Francesco Vinci\inst{1} \and
Gyunam Park\inst{2} \and 
Wil M. P. van der Aalst\inst{2,3}
 \and
Massimiliano~de~Leoni\inst{1}
}
\authorrunning{F. Vinci et al.}
\institute{University of Padua, Italy \\
\email{francesco.vinci.1@phd.unipd.it}, 
\email{deleoni@math.unipd.it}\and
Fraunhofer Institute for Applied Information Technology (FIT),  Germany \\
\email{gyunam.park@fit.fraunhofer.de} \and
Process and Data Science, RWTH Aachen University, Aachen, Germany \\
\email{wvdaalst@pads.rwth-aachen.de}
}
\maketitle              % typeset the header of the contribution
\begin{abstract}
Business Process Simulation (BPS) refers to techniques designed to replicate the dynamic behavior of a business process. 
Many approaches have been proposed to automatically discover simulation models from historical event logs, reducing the cost and time to manually design them.
However, in dynamic business environments, organizations continuously refine their processes to enhance efficiency, reduce costs, and improve customer satisfaction. 
Existing techniques to process simulation discovery lack adaptability to real-time operational changes.
In this paper, we propose a streaming process simulation discovery technique that integrates Incremental Process Discovery with Online Machine Learning methods. This technique prioritizes recent data while preserving historical information, ensuring adaptation to evolving process dynamics.
Experiments conducted on four different event logs demonstrate the importance in simulation of giving more weight to recent data while retaining historical knowledge. Our technique not only produces more stable simulations but also exhibits robustness in handling concept drift, as highlighted in one of the use cases.

% Conventional process simulation discovery techniques rely on historical event logs, providing valuable insights but lacking adaptability to real-time operational changes. 
% In this paper we propose a streaming process simulation discovery approach that integrates Incremental Process Discovery with Online Machine Learning techniques. This approach prioritizes recent data while preserving historical information, ensuring adaptation to evolving process dynamics.
% Experiments conducted on four different case studies demonstrate the importance in simulation of giving more weight to recent data while retaining historical knowledge. Our method not only produces more stable simulations but also exhibits robustness in handling concept drift, as highlighted in one of the case studies.

\end{abstract}
\section{Introduction}
\label{sec:intro}
Business Process Simulation (BPS) is one of the most used techniques for analyzing and improving business processes. By incorporating key aspects, such as activities control-flow, task durations, and resource allocation, BPS can capture a probabilistic characterization of various run-time aspects. Then, they enable organizations to evaluate different scenarios, anticipate bottlenecks, and make data-driven decisions. BPS models can be manually designed, requiring extensive domain knowledge and significant effort. To overcome these limitations, automated approaches leveraging historical event logs have been developed, enabling the complete discovery of simulation models (cf.\ Section~\ref{sec:rel_works}).

However, dealing with evolving processes and with their dynamic behavior remains one of the major challenges. Organizations continuously adapt their processes in response to internal policy changes or external factors. Traditional techniques for discovery of
simulation models fail to capture these ongoing changes because they do not feature possibilities to update the discovered models as changes are observed in the process' behavior. This affects the process when a sudden concept drift occurs, as the event log treats both older and newer behavior with equal importance in process simulation model discovery. As a result, the discovered model would simulate behavior from both before and after the drift as if it were future behavior, likely leading to invalid conclusions.

In this paper, we propose a novel \textbf{streaming process simulation discovery} technique that integrates Incremental Process Discovery~\cite{danielThesis} with Online Machine Learning methods. Our technique continuously updates the simulation model by incorporating new event behaviors while preserving historical knowledge. Specifically, we employ Hoeffding Adaptive Trees~\cite{hoeffding}, which are well-suited for evolving data streams and can dynamically adapt to process changes over time. By prioritizing recent data without discarding valuable past information, our technique enhances the accuracy and stability of simulation models.

% The evaluation is based on four different processes and corresponding event logs.
% The event logs temporal dimensions were divided into $n$ temporal windows $W_1, \ldots, W_{n}$, with $t_i$ being the latest timestamp of $W_i$.
% % The event logs were divided into $n$ batches by counting the number of weeks from the earliest to the latest timestamp and grouping them into batches $W_1, \ldots, W_{n}$, with $t_i$ being the latest timestamp in batch $W_i$.
% Process simulation models were discovered for all $t_1,\ldots,t_{n}$ using (a) all traces contained in $W_1 \ldots, W_i$, or (b) the traces that belong to $W_i$, or (c) using our incremental technique on all windows $W_1 \ldots, W_i$.
% The three simulation models at any $t_i$ - one per technique - were used to generate synthetic event logs, whose traces were compared with those starting in $W_{i+1}$, using consolidated, literature metrics.
%\color{blue}
The evaluation compares our streaming discovery technique with two baselines. A first baseline considers all historical event data, including those before any process' behavioral drifts; a second only uses the recent data, which would not mix process variants with different behavior.
%\color{black}
% ~\autoref{fig:train_test} illustrate the evaluation methodology, comparing the three techniques, where (a) and (b) serve as baseline reference for evaluation. 
This comparison highlights the effectiveness of our technique and determines when all historical data are necessary. The results show that our technique produces more reliable simulations and effectively adapts to evolving processes. 

The rest of the paper is structured as follows: Section~\ref{sec:motivation} illustrates a motivational example, Section~\ref{sec:rel_works} reviews related works on process simulation and online learning techniques. Section~\ref{sec:method} details our proposed technique, outlining how it integrates business process simulation discovery with adaptive learning. Section~\ref{sec:exp} presents the experimental setup and evaluation results. Finally, Section~\ref{sec:conclusion} concludes the paper.

\section{Motivating Example}
\label{sec:motivation}
\begin{figure}[t!]
\centering
\begin{subfigure}{.8\textwidth}
  \centering
  \includegraphics[width=\linewidth]{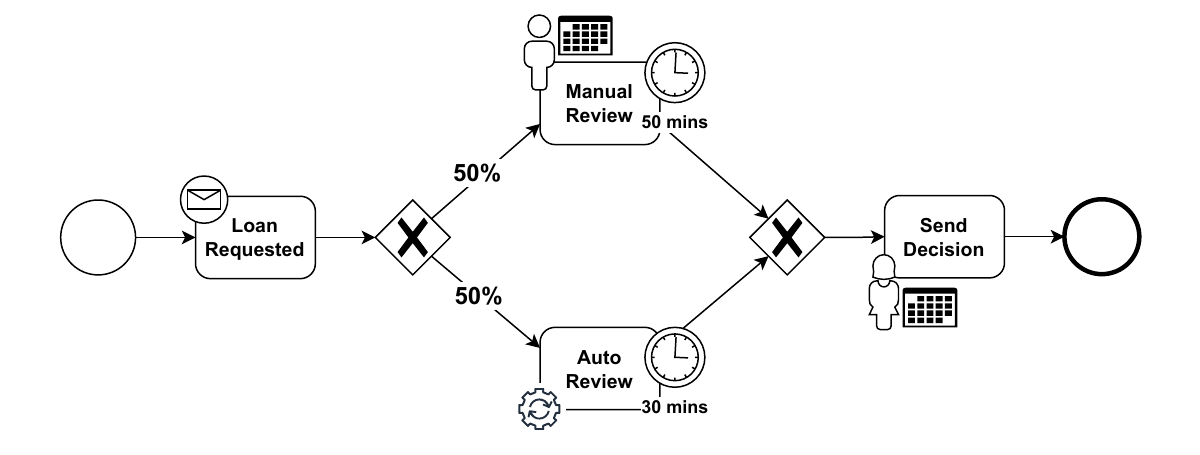}
  \caption{Simulation model before the concept-drift.}
  \label{fig:model_example_0}
\end{subfigure}
\begin{subfigure}{.8\textwidth}
  \centering
  \includegraphics[width=\linewidth]{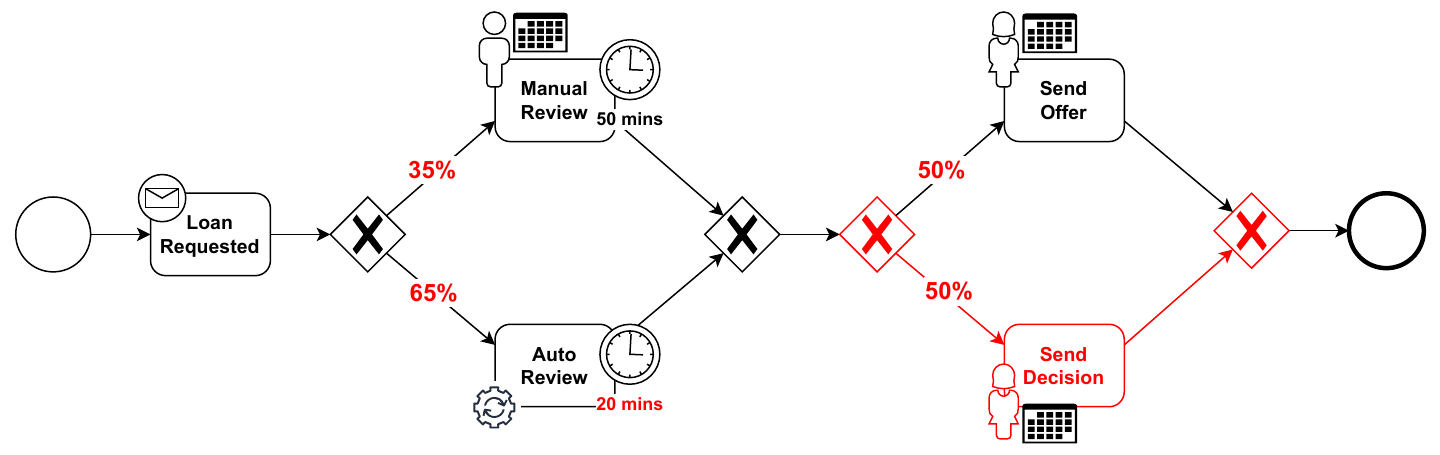}
  \caption{Simulation model discovered after the concept-drift by using traditional process simulation discovery approaches. Red parts denote inaccuracies.}
  \label{fig:not_accurate_model_example}
\end{subfigure}
\begin{subfigure}{.8\textwidth}
  \centering
  \includegraphics[width=\linewidth]{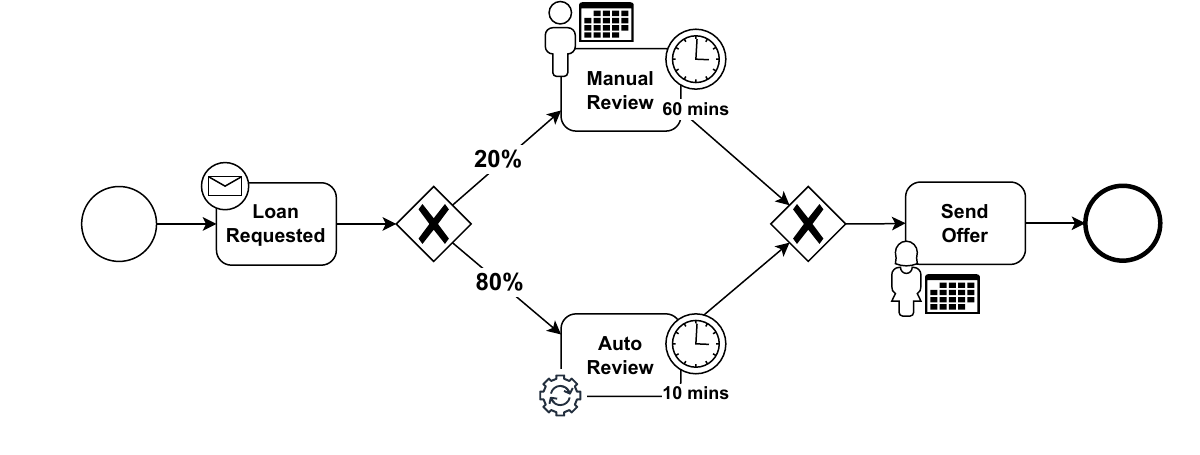}
  \caption{Simulation model discovered after the concept-drift by using online process simulation discovery techniques.}
  \label{fig:new_model_example}
\end{subfigure}
\caption{Loan application process model examples. The control-flow process models is illustrated using BPMN notation, where "X" represents a decision point. Percentages in the arcs represent path probabilities after the decision points. Clock indicates activities durations in minutes. Resources, where involved, are represented with icons and calendars.}
\label{fig:example_model}
\end{figure}

In this section, we present an example that highlights the motivation for proposing an online process simulation discovery technique.
Consider a loan application process within a financial institution. Initially, when a customer submits a \textit{loan request}, the application is processed through one of two paths: either a \textit{manual review} by an expert ($50\%$ of probability, with $50$ minutes duration) or an \textit{automated approval} by a system ($50\%$ of probability, with $30$ minutes duration).
The process concludes with the \textit{notification of the decision}. \autoref{fig:model_example_0} depicts a simulation model that a process simulation discovery technique would generate based on historical data. 

However, due to internal decisions to optimize the process, the institute decided to enhance its automated system, increasing the probability of automated reviews to $80\%$ while reducing its processing time to $10$ minutes. Moreover, rather than simply concluding with an approval or rejection, the process outcome now includes a \textit{loan offer}.

When discovering simulation models using traditional process simulation discovery approaches, we discover the model depicted in \autoref{fig:not_accurate_model_example}.
As we incorporate all past event data (i.e., with pre- and post-drift behaviors), we fail to accurately reflect the changes.
Instead, if we properly prioritize the recent data, we can discover the simulation model described in \autoref{fig:new_model_example}. At the same time, previous data information are crucial for capturing additional observations, such as resource calendars and other event attributes.
In this paper, we propose a technique for incrementally updating previously discovered simulation models, ensuring they adapt to process changes while maintaining high accuracy.

% After this process drift, the process model must be updated, and process simulation discovery techniques can be re-applied. However, incorporating all past event data would result in a process model that averages pre- and post-drift behaviors, leading to inaccurate simulation models that fail to accurately reflect the changes. In \autoref{fig:not_accurate_model_example} we illustrate an example of such process model, highlighting in red the inaccurate parts.

% The goal of this paper is to propose a technique for incrementally updating previously discovered simulation models, ensuring they adapt to process changes while maintaining high accuracy.

\section{Related Work}
\label{sec:rel_works}
The work by Rozinat et al.~\cite{ROZINAT2009305} is one of the first combining Process Mining techniques to discover multiple perspectives of a process (control-flow, data, performance, and resource aspects) and integrating them into complete simulation models.
In~\cite{CAMARGO2020113284}, Camargo et al. presented a method to discover business process simulation models from event log data with the goal to optimize the accuracy of it. 
The increasing availability of data and the advancement of new Machine and Deep Learning techniques led to the integration of these into traditional Business Process Simulation methods. 
Camargo et al.~\cite{CAMARGO2023102248} and Meneghello et al.~\cite{MENEGHELLO2025102472} propose two hybrid approaches where a process model is discovered to model the process' control-flow perspective, which is extended with Deep Learning models for the run-time characterization of the other perspectives. While these studies primarily aimed to enhance the accuracy of temporal modeling through Machine Learning, de Leoni et al.~\cite{deLeoniBPM2023} focuses on improving control-flow accuracy by incorporating logistic regression models into the process model.

These discovery techniques aim to generate a process simulation model from an input event log. However, they assume that the process remains stable over time, analyzing patterns by averaging past and recent behaviors. In reality, processes may be dynamic and evolve over time due to external factors or internal process optimizations. This phenomenon, known as concept drift, occurs when the process behavior changes over time, potentially making previously discovered models inaccurate. 
%\color{blue}
To address this, several works proposed approaches for detecting, localizing and dealing concept drifts~\cite{Bose,Sato2021}. Recent studies have introduced techniques and methodologies for detecting concept drift across multiple process perspectives—including control-flow, resources, and performance—showing how processes can evolve in complex and multifaceted ways~\cite{Adams2021,Klijn2024,Kraus2025}. 

Other works proposed approaches for detecting concept drifts from event streams~\cite{Lu2021}.
Process mining techniques applied to the analysis of data streams are referred as \textit{streaming process mining}~\cite{Burattin2022}. These techniques can be used to dynamically updating existing process models~\cite{Burattin2014,Carmona2012,Bas2018}.
Navarin et al.~\cite{Navarin2020} presented a technique for discovering declarative process models from event streams that incorporates both control-flow dependencies and data conditions. In~\cite{Scheibel2023} Scheibel and Rinderle-Ma presented an online decision mining technique using an adaptive window technique (ADWIN)~\cite{hoeffding} to detect changes. However, these works have primarily focused on the control-flow and decision mining perspectives, without incorporating multiple perspectives essential for BPS models. 

In the domain of predictive process monitoring, Rizzi et al.~\cite{Rizzi_2022} evaluated three different strategies that support either the periodic rediscovery or the incremental construction of predictive models, thereby allowing models to stay up-to-date with new data. However, these approaches are not specifically designed for the discovery or updating of simulation models.
%\color{black}

\section{Online Process Simulation Discovery}
\label{sec:method}
This section introduces a novel technique for discovering Business Process Simulation models from event data, designed to remain robust despite changes over time.
We firstly define event streams and Business Process Simulation models, and then present our discovery technique.

Typically, BPS models consist of a process model represented as graph (e.g., Petri nets, BPMN models, or process trees) to capture activity control-flow~\cite{vanderAalst2016}, enhanced with additional parameters representing various perspectives, such as time, resources, and data attributes. State-of-the-art methods have introduced hybrid simulation models that integrate these multi-perspective parameters using Machine Learning techniques, showing their potential in improving overall performances~\cite{CAMARGO2023102248}.

Our technique combines \textbf{Incremental Process Discovery} with \textbf{Online Machine Learning} to continuously update hybrid BPS models. The general idea is to start with an initial process simulation model and progressively refine it using streaming data. The rationale is that we aim to give more weight to most recent data, then adapting the simulation model to current trends while still retaining valuable insights from past data.

In the following sections, we first introduce key preliminary concepts, including streaming of events and business process simulation models (cf. Section~\ref{sec:preliminaries}). These concepts provide the foundation for describing our technique in Section~\ref{sec:approach}.

\subsection{Preliminaries}
\label{sec:preliminaries}
Process data are typically collected as sequences of events, defining traces and event logs~\cite{vanderAalst2016}. Let $\ME$ be the universe of events. Given an event $e\in\ME$ we assume the following projections: $case(e)\in\MI$ the case identifier, $act(e)\in\MA$ the activity executed, $res(e)\in\MR$ the resource involved, $time(e)\in\mathbb{N}$ the timestamp, and $attr(e)\in\MV$ a vector of event attributes, where $\MI$, $\MA$, $\MR$ and $\MV$ represent the sets of all possible case identifiers, activities, resources and event attributes, respectively.
A trace is defined as a sequence of events ordered by timestamp and sharing the same case identifier. An event log is a set of such traces.
% In this paper, we assume that an event is always associated with a process case, also referred to as a trace, and contains information such as the activity performed, the resource involved, times, and other attributes.
% \begin{definition}[Events]\label{def:events}
%     Let $\MI$ be the set of case identifiers, $\MA$ the set of process activity labels, $\MR$ the set of process resources, $\MT\subseteq \mathbb{N}$ the set of process timestamps, $\MV$ a set of additional attributes. 
%     An event is a tuple $(c,a,r,t, v)\in\MI\times\MA\times\MR\times\MT\times\MV$, where $c$ is the case identifier, $r$ the resource performing the activity $a$ at a timestamp $t$, and $v$ a vector of event attributes.
% \end{definition}
% In the reminder, given an event $e=(c,a,r,t,v)$ we use define $case(e)=c$, $act(e)=a$, $res(e)=r$, $time(e)=t$ and $attr(e)=v$. Sequence of events define traces and event logs, which are the core ingredient for describing a process. 

% \begin{definition}[Traces \& Event Logs]\label{def:eventlog}
%     Let $\ME=\MI\times\MA\times\MR\times\MT\times\MV$ be the universe of events. A trace $\sigma=\langle e_1,\dots,e_n\rangle\in\ME^*$ is a sequence of events ordered by timestamp and sharing the same case identifier, i.e. $case(e_i)=c\ \forall j=1,\dots,n$ and $time(e_i)\leq time(e_{i-1})\ \forall i=2,\dots,n$. An event log $\ML$ is a set of traces.
% \end{definition}

Process Mining techniques aim to create models and discover process patterns analyzing event logs~\cite{vanderAalst2016}. Traditionally, these techniques take as input an entire set of completed traces and use it for deriving conclusion. In this paper, we assume to deal with \textbf{event streams}, i.e. a sequence of unique events~\cite{basThesis}.
\begin{definition}[Event Stream]\label{def:streaming}
    Let $\ME$ be the universe of events. An event stream $\MS=\langle\dots,e_i,e_{i+1},\dots\rangle\in\ME^*$ is an infinite sequence of events such that, for any $i\ge 1$, $time(e_{i})\leq time(e_{i+1})$.
    % Given a time window $w>0$, we define $\MS_{w,t}^{\ME}=\left\{e\in\MS_t^{\ME}\mid time(e)\in\left[t-w, t\right]\right\}$ as the subset of events occurring within the most recent time window of length $w$.
\end{definition}
% \begin{definition}[Streaming of Events]\label{def:streaming}
%     Let $\ME=\MI\times\MA\times\MR\times\MT\times\MV$ be the universe of events. A streaming of events at a timestamp $t\in\MT$ is a sequence of events $\MS_t^{\ME}=\langle e_1,\dots,e_k\rangle\in\ME^*$ such that $time(e_{i-1})\leq time(e_i)\leq t\ \forall i=2,\dots,k$.
%     Given a time window $w>0$, we define $\MS_{w,t}^{\ME}=\left\{e\in\MS_t^{\ME}\mid time(e)\in\left[t-w, t\right]\right\}$ as the subset of events occurring within the most recent time window of length $w$.
% \end{definition}
Streaming process data enable the adaptive discovery of process models. This paper specifically focuses on the incremental learning of \textbf{Business Process Simulation models}. We leverage on hybrid process simulation models, which integrate Process Mining and Machine Learning techniques, resulting in more accurate simulation models~\cite{CAMARGO2023102248,deLeoniBPM2023,MENEGHELLO2025102472}. 

Formally, a Business Process Simulation (BPS) model is defined as a tuple $M=(N, D, P)$ where
     $N$ is the activity control-flow process model,
     $D$ is the set of descriptive parameters, and
     $P$ the set of predictive parameters that characterize temporal and stochastic perspectives. 

Specifically, $N$ can be represented as a Petri net model, BPMN diagram or process tree~\cite{vanderAalst2016}. The descriptive parameters in $D$ include the set of resources and what activities they perform, their working calendars, and the event attribute distributions. The predictive set $P$ consists of models for estimating execution times of activities, resource waiting times, and case arrival rates. Additionally, it includes predictive models for determining branching probabilities~\cite{deLeoniBPM2023}. In this paper, we assume all these models to be probabilistic decision trees, ensuring both explainability and the stochastic nature essential for simulation. The predictive models can then be integrated at runtime during the simulation to generate complete simulated event logs~\cite{MENEGHELLO2025102472}.

These BPS models can be discovered from event data by combining Process Mining techniques to obtain the process model and descriptive parameters, with Machine Learning algorithms to train the predictive models~\cite{CAMARGO2023102248}. 

This formulation enables the application of Incremental Process Discovery techniques to dynamically refine the process model $N$, while Online Machine Learning methods can be applied to continuously update the predictive parameters $P$, defined as Machine Learning models.

%\color{blue}
In this work, we employed Hoeffding Adaptive Trees (HATs)~\cite{hoeffding} as the core online learning method for updating the predictive models in $P$. Like the original Hoeffding Tree, HAT leverages the Hoeffding bound to make statistically sound decisions about node splits based on a limited number of examples, ensuring efficient, incremental learning. The adaptive component of HAT addresses concept drift by incorporating an adaptive windowing method (ADWIN) that monitors performance at each node. When a drift is detected, HAT can replace underperforming branches with alternate subtrees that better capture the current concept. This permits the predictive models to evolve continuously, ensuring the simulation model remains accurate over time.
%\color{black}

\subsection{Our Discovery Technique}
\label{sec:approach}
Traditional process discovery techniques rely on analyzing historical event data in a single batch to derive a process model. These methods assume that the process remains static over time, resulting in models that may become outdated as the process evolves dynamically.

In this paper, we propose a technique for incrementally discovering process simulation models. The starting point of the technique is a simulation model $M_{t_0}=(N_{t_0}, D_{t_0}, P_{t_0})$, which represents a process based on information available up to timestamp $t_0$. Given a new set of events at timestamp $t>t_0$, the goal is to produce an updated simulation model $M_t=(N_t, D_t, P_t)$ integrating newly observed behaviors while refining the existing model.

We formalize this problem by defining a function $\Phi:\MM\times \ME^*\to\MM$, where $\ME$ is the universe of possible events, and $\MM$ denotes the universe of all possible process simulation models. Given an existing process simulation model $M\in\MM$ and a new sequence of events $\MS\in\ME^*$, the function returns an updated simulation model $\Phi\left(M, \MS\right) = M'\in\MM$ able to replicate new observations.

The procedure for updating the existing simulation model can be divided into four key steps. First, data preparation is performed to select and structure the event sequence for model updating. Next, the control-flow model is refined using Incremental Process Discovery techniques. 
Then, descriptive parameters are updated. 
Finally, the predictive parameters (models) are adjusted using Online Machine Learning techniques. 
% Given a timestamp $t$, the updated simulation model $M_t$ is obtained through the following steps.

\subsubsection{Step 0: Event Data Preparation}
\label{sec:step0}
\begin{figure}[t!]
    \centering
    \includegraphics[width=0.59\linewidth]{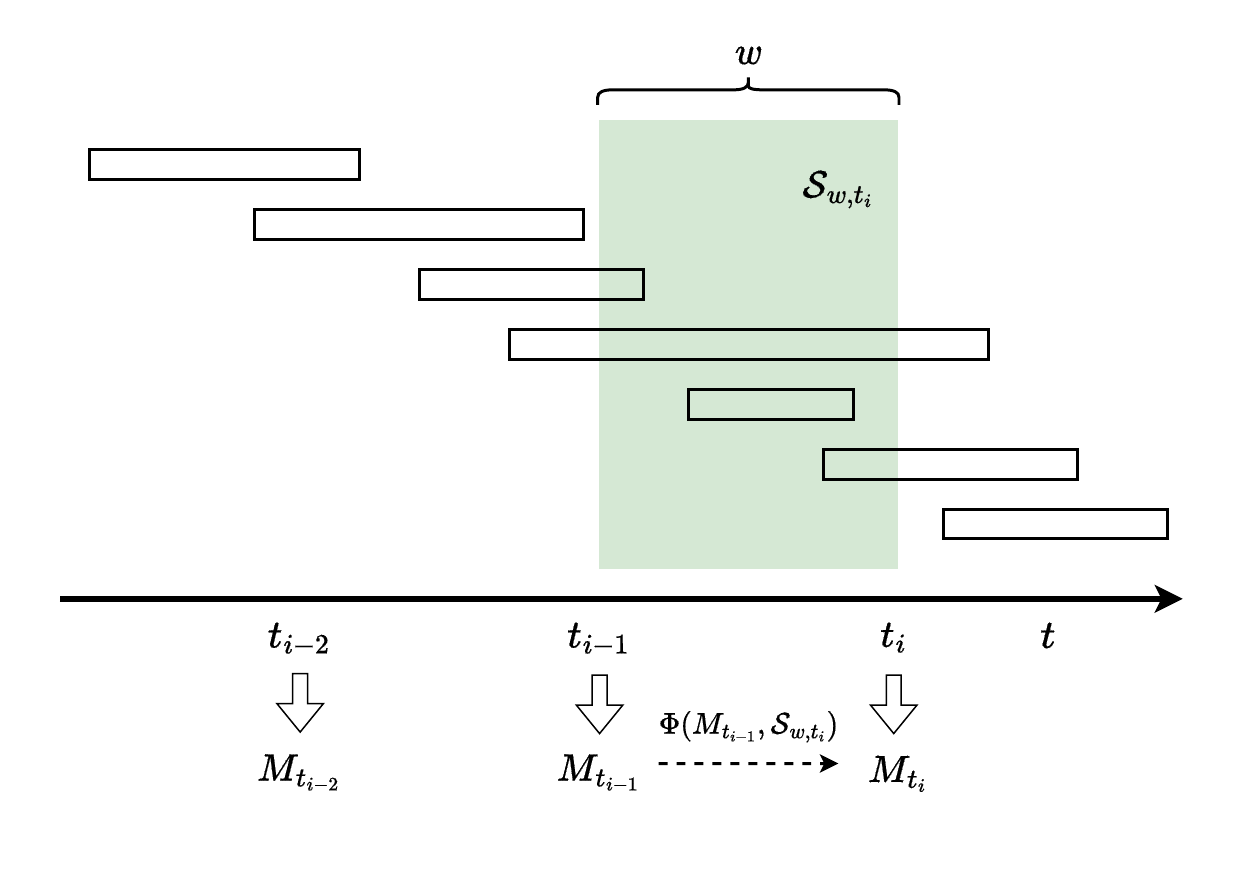}
    \caption{Iterative procedure for online process simulation discovery at each timestamp $t_i$ using window size $w$. Rectangles represent traces over time. The green part represents the time window containing the events in $\MS_{w,t_i}$ used for incrementally updating the simulation model at time $t_i$.}
    \label{fig:online_window}
\end{figure}

Conventional Process Mining techniques are designed to operate on sets of traces as input. Several techniques have been proposed to extract finite sets of events from event streams for use in Process Mining~\cite{basThesis}. In this paper, we adopt the concept of \textbf{sliding window}, but note that the technique is generalizable to any other extraction approaches.
\begin{definition}[Sliding Window]
    Let $\MS$ an event stream. 
    Given a timestamp $t\in\mathbb{N}$, and the window size $w\in\mathbb{N}^{>0}$.
    We define the sliding window of size $w$ at timestamp $t$ as 
    $
    \MS_{w,t}=\langle e\in\MS\mid time(e)\in\left[t-w,t\right]\rangle
    $
\end{definition}
The rationale is that a process analyst would update the simulation model at regular time intervals defined by a predefined window size $w$. This means that a new model is generated every $w$ time units. For example, if $w$ is set to one week, the simulation model is updated weekly to reflect the latest observed process changes during that week. 
This periodic update mechanism can be formalized as iteratively refining an initial process simulation model $M_{t_0}$ at a timestamp $t_0$. The updated process simulation model at a timestamp $t$ is given by $M_t=M_{t_i}=\Phi\left(M_{t_{i-1}}, \MS_{w,t_i}\right)$, where $i>0$, $t_{i}\leq t<t_{i+1}$, $t_i=t_{i-1}+w$ and $\MS_{w,t_i}$ represents the stream of events occurred within the most recent time window of length $w$.  \autoref{fig:online_window} illustrates this discussion, showing how the simulation models are obtained over time for a time window of size $w$. This approach defines a sequence of simulation models, each corresponding to a specific time window, where one extends the previous.

\subsubsection{Step 1: Control-Flow Model Update}
\label{sec:step1}
The previous process model $N_{t_{i-1}}$ is incrementally updated to incorporate new observed behaviors from the last observed data in the sliding window $\MS_{w,t_i}\in\ME^*$. We employ Incremental Process Discovery techniques to refine the existing process model, resulting in an updated model $N_{t_i}$ able to capture the new observations~\cite{danielThesis}.  $\MS_{w,t_i}$ may include incomplete traces - prefixes, infixes or postfixes - in addition to complete traces (see \autoref{fig:online_window}). To effectively handle this, we build upon the work by Schuster et al.~\cite{IPD_trace_fragments} designed for applying an Incremental Process Discover technique using trace fragments.
Trace fragments can be easily reconstructed from the stream of events by projecting on the case identifiers.
%\color{blue}
Specifically, prefixes refer to cases that start within the window but are not yet completed, postfixes to cases that complete in the window but started earlier, and infixes are fragments of ongoing cases that both started before and continue beyond the current window. We assume that domain knowledge can be used to determine when cases reach a completion. This knowledge can be given in form of possible activities that mark the process executions, or alternatively by setting a timeout, namely a case is assumed to be completed if no new activity is observed within a time threshold~\cite{Bas2018}. 

We acknowledge that this can be a threat of validity in certain settings. However, this is plausible in other settings. For example, in a loan application process, a case ends when the offer is either accepted or rejected; in a purchase process, upon completion of payment; and in a pharmacy retail setting, when the prescription is fulfilled (see Section~\ref{sec:setup}).
%Then traces and fragments that are not replicable in the current process model are selected as input for the Incremental Process Discovery technique, which updates the previous process model $N_{t_{i-1}}$ and produces a refined model $N_{t_i}$ capable of replicating the trace (fragments) in $\MS_{w,t_i}$.
%\color{black}
% Then, these can be used as input for the Incremental Process Discovery technique, which updates the previous process model $N_{t_{i-1}}$ and produces a refined model $N_{t_i}$ capable of replicating the trace fragments in $\MS_{w,t_i}$. 

Notably, the method by Schuster et al.~\cite{IPD_trace_fragments} assumes that the input process model is a process tree. Process trees can be easily converted into Petri nets or BPMN diagrams and vice versa under the assumption that they are block structured~\cite{vanderAalst2016}. However, this does not pose a limitation to our technique, since many process discovery algorithms, such as Inductive Miner~\cite{Leemans13}, produce block-structured models, which ensure soundness and validity.

\subsubsection{Step 2: Descriptive Parameters Update}
\label{sec:step2}
The set of descriptive parameters $D_{t_{i-1}}$ is updated by including the new behaviors in the sliding window $\MS_{w,t_i}$. Specifically, when previously unobserved resources appear in $\MS_{w,t_i}$, they are integrated in the set of parameters with their associated working schedules and the activities they perform. Moreover, for resources already present in $D_{t_{i-1}}$, their calendars are recomputed in the new sequence of events $\MS_{w,t_i}$. Similarly, the event data distribution are adjusted computing them using the events in $\MS_{w,t_i}$.
This results in an updated set of parameters $D_{t_i}$.

\subsubsection{Step 3: Predictive Models Update}
\label{sec:step3}
Finally, the predictive models in $P_{t_{i-1}}$ are updated to reflect changes in the process observed within the sliding window $\MS_{w,t_i}$. These updates occur in two ways:
\begin{itemize}
    \item The updated process model $N_{t_i}$ and parameter set $D_{t_i}$ may introduce previously unseen elements, such as new resources, new activities, or new pathways in $N_{t_{i}}$ that were not present before. 
    In such cases, new predictive models are defined. First, for any new activity, we train a new probabilistic decision tree to estimate its execution time. 
    Second, for any new resource, we train a new probabilistic decision tree to estimate the waiting times. Finally, if a new pathway is introduced in the process model, decision trees are defined to determine the probability of selecting that path. These new models are then included in the new set of predictive parameters $P_{t_i}$.
    \item Existing predictive models in $P_{t_{i-1}}$ are continuously refined using Online Machine Learning techniques. These methods enable incremental updates, allowing models to evolve with the latest observed behaviors without discarding previous knowledge. Several Online Machine Learning techniques exist in the literature; however, we chose Hoeffding Adaptive Trees~\cite{hoeffding} due to their explainability and ability to adaptively learn from data streams that evolve over time, making them robust to concept drift.
    These trees incrementally update their structure based on incoming data, using statistical tests to determine when to split nodes, enabling efficient and adaptive learning. 
    Moreover, Hoeffding Adaptive Trees incorporate drift detection mechanisms, selectively updating branches of the tree when significant changes in the data distribution are detected. 
\end{itemize}
Through these updates, the set of predictive models evolves into an updated set $P_{t_i}$, contributing to the refinement of the overall simulation model.
The final resulting simulation model $M_t=M_{t_i}=(N_{t_i},D_{t_i},P_{t_i})$ incrementally incorporates the behaviors observed in the most recent data window $\MS_{w,t_i}$, thus being representative for the upcoming period. 
% This ensure that the simulation remains an accurate and up-to-date representation of the evolving process, capable of making reliable predictions and analyses.

\section{Experiments}
\label{sec:exp}
In this section, we present the experiments conducted to evaluate the performance and applicability of our discovery technique. The goal is to show its ability to produce accurate process simulations over time. To this aim, we monitored the accuracy by dividing the temporal dimension into multiple windows and computing accuracy metrics for each. The evaluation focuses on the distances proposed by Chapela-Campa et al. in~\cite{CHAPELACAMPA2025102447}, which assess the accuracy of the simulation discovery technique from various perspectives. The metrics are based on computing distances between the actual event log and simulated ones.

We compared our results using three different discovery techniques: (a) the \textbf{single large batch} technique, which uses all previous data, (b) the \textbf{last small batch} technique, which considers only the data from the most recent time window, and (c) our proposed \textbf{online} discovery technique. Specifically, (a) and (b) serve as baseline techniques for comparison. This evaluation aims to assess the importance of prioritizing the most recent data (online/last vs single batch), and preserving historical information (online/single vs last batch) (cf. Section~\ref{sec:setup}).

\subsection{Experimental Setup}
\label{sec:setup}
We implemented our discovery technique in Python, where the simulator is initialized with a process model represented as a Petri net~\cite{vanderAalst2016}, and a set of parameters that can be discovered from an event log.\footnote{\url{https://github.com/franvinci/ProcessSimulationTool}} Once an initial simulation model is defined, it can be further enhanced through continuous event streaming.
We used the implementation of~\cite{IPD_trace_fragments} in the Cortado library~\cite{cortado}. Additionally, we integrated the River library~\cite{river} which implements Hoeffding Adaptive Trees.

We conducted experiments on four different event logs, each representing a different process:
\begin{description}
    \item[BPIC17W.] It is the subprocess for the workflow-relevant activities, i.e., those starting with W, in the BPI Challenge 2017 event data, a log of a loan application process from a Dutch financial institute.\footnote{\url{https://doi.org/10.4121/uuid:5f3067df-f10b-45da-b98b-86ae4c7a310b}}
    \item[BPIC12W.] It is the same subprocess as BPIC17W but referring to the 2011 process executions, which are recorded in the BPI Challenge 2012 data.\footnote{\url{https://doi.org/10.4121/uuid:3926db30-f712-4394-aebc-75976070e91f}}
    \item[Purchase to Pay (P2P).] It is a realistic purchasing example process with synthetic event log.\footnote{\url{https://fluxicon.com/academic/material}}
    \item[CVS retail pharmacy (CVS).] It refers to a realistic pharmacy retail process with synthetic event log.\footnote{\url{https://zenodo.org/records/4699983}}
\end{description}
The temporal dimensions of the event logs have been divided into $10$ temporal windows as follows: we counted the number of weeks from the earliest to the latest timestamp, and split the weeks in 10 obtaining the windows $W_1, \ldots, W_{10}$. 
%\color{blue} 
The choice of 10 windows aims to balance the significance of each window, on the one hand, and the satisfactory frequency of model updates. Using more than 10 windows would reduce the data per window to less than 10\% of the total, potentially compromising the statistical reliability of each update. On the other hand, using fewer than 10 windows would result in infrequent model updates.

% We selected a temporal granularity of 10 windows, as this provided a balanced and meaningful segmentation of the event log. Each window accounts for $10\%$ of the total time span and is used to test the model trained on the preceding data, ensuring that the evaluation is conducted over a substantial and representative portion of the data. Increasing the number of windows would result in smaller test sets, potentially reducing the reliability of accuracy measurements. Conversely, reducing the number of windows would lead to fewer model updates, limiting the depth of insights into the simulation model evolution. This design choice reflects a practical trade-off and does not compromise the validity of the method, which remains general and robust across different temporal granularities.
%\color{black}
% The event logs have been divided into $10$ batches as follows: we counted the number of weeks from the earliest to the latest timestamp, and split the weeks in 10 groups $G_1, \ldots, G_{10}$. For each group $G_i$, we obtain a batch $W_i$ that contains every event that ends in the weeks in $G_i$.
Denote the latest timestamp in $W_i$ with $t_i$, we consider process simulation models at timestamps $t_1,\ldots,t_{10}$, which we discovered via two baseline techniques and ours. In particular, the simulation model at timestamp $t_i$ is discovered as follows for two baseline techniques, namely technique (a) and (b), and our technique proposal, i.e.\ technique (c):
\begin{enumerate}[label=(\alph*)]
    \item We employed traditional - not incremental - discovery techniques for process simulation models (see below), using the traces contained in $W_1,\ldots,W_i$.
    \item We employed traditional discovery techniques for process simulation models, using the traces in the only window $W_i$.
    \item Our online discovery technique began by creating an initial simulation model using completed traces from the first window $W_1$.  The simulation model is  incrementally updated using events in $W_2, \ldots, W_i$.
\end{enumerate}
\begin{figure}[t!]
\centering
\begin{subfigure}{.33\textwidth}
  \centering
  \includegraphics[width=\linewidth]{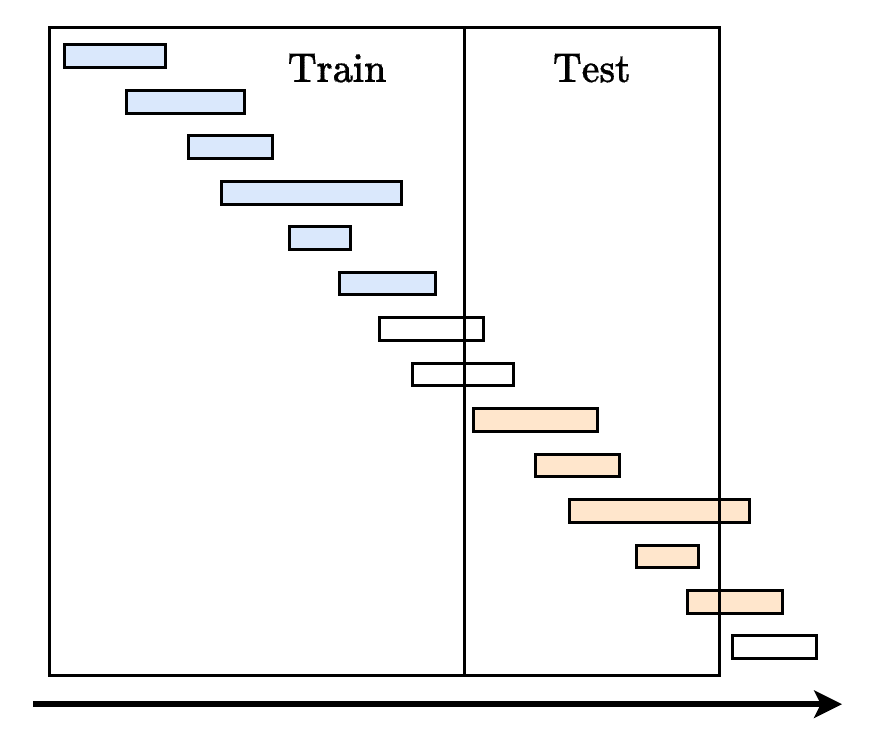}
  \caption{Single large batch.}
  \label{fig:train_onebatch}
\end{subfigure}%
\begin{subfigure}{.33\textwidth}
  \centering
  \includegraphics[width=\linewidth]{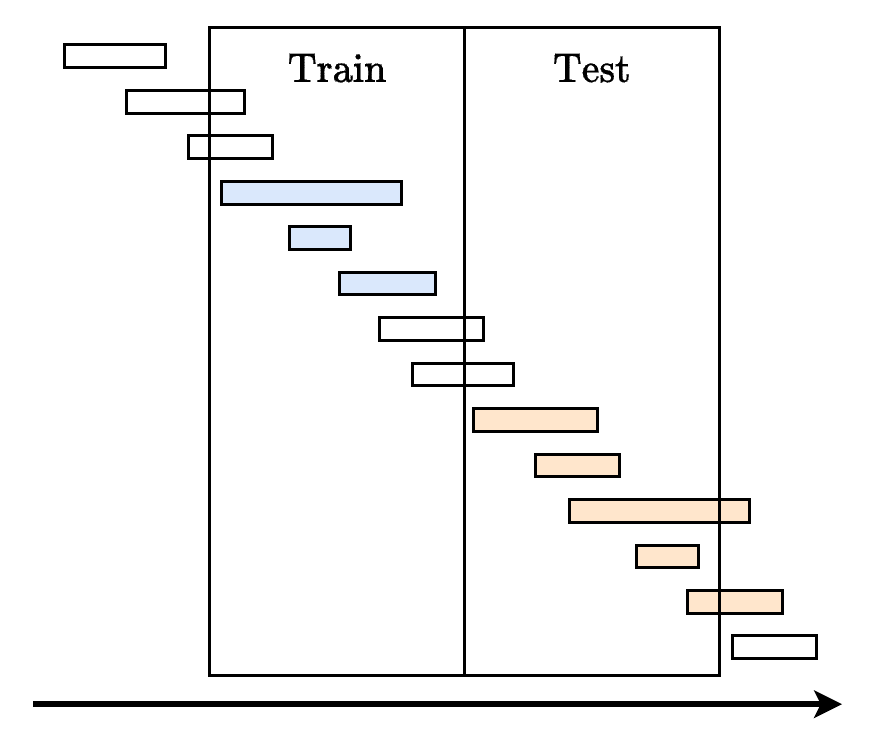}
  \caption{Last small batch.}
  \label{fig:train_lastbatch}
\end{subfigure}
\begin{subfigure}{.33\textwidth}
  \centering
  \includegraphics[width=\linewidth]{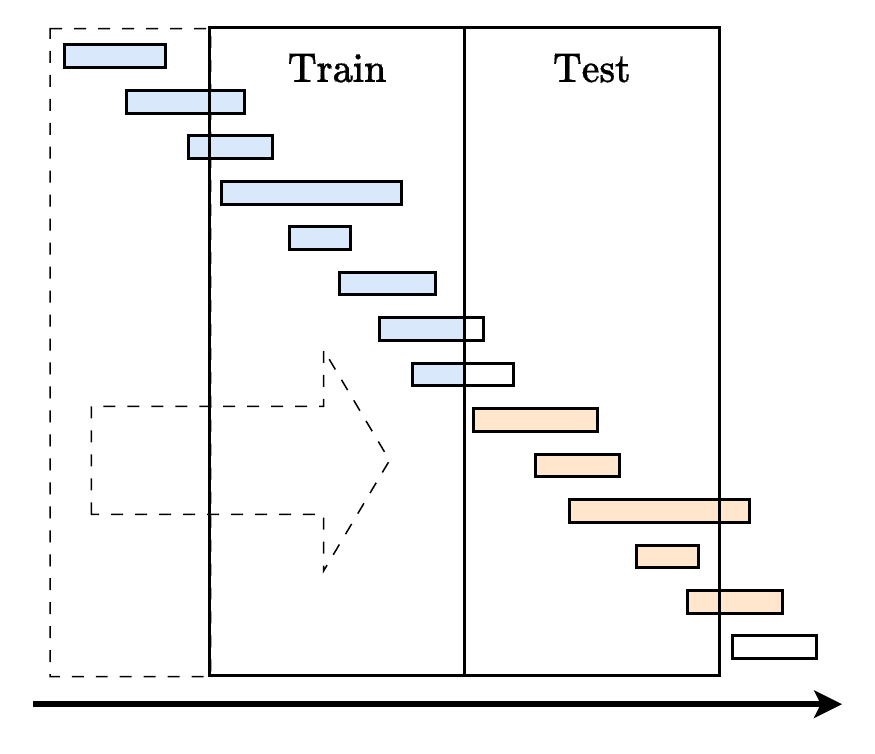}
  \caption{Online.}
  \label{fig:train_online}
\end{subfigure}
\caption{Illustration of the evaluation methodology. Rectangles depict traces over time. The blue color indicates the events used for training the simulation model. The orange indicates the traces used for testing it.}
\label{fig:train_test}
\end{figure}
The experimental results at timestamp $t_i$ are those measured against the test set containing completed traces that started in the following window, i.e. $W_{i+1}$. 
Specifically, the obtained simulation model at time $t_i$ is used for generating an event log that is then compared with the event log containing traces started in window $W_{i+1}$.
Notably, for the P2P process, no trace was in $W_{10}$, preventing the assessment of the model obtained at timestamp $t_9$, which is thus not considered in the results. The same has happened for the CVS process where four windows $W_7,\ldots,W_{10}$ were empty.
\autoref{fig:train_test} illustrates the three cases. Note that compared to other approaches, our online technique allows the use of trace fragments. This comparison examines whether assigning greater weight to recent data improves the accuracy of simulation models for predicting the future. At the same time, we also explore whether retaining historical data is essential for better estimating simulation parameters.

For the techniques (a) and (b) in the list above, control-flow process models have been discovered using Inductive Miner~\cite{Leemans13}, while the decision trees modeling the predictive models are obtained via the CART algorithm, and imposing a maximum depth of $5$ for maintaining explainability and avoid overfitting.

For our technique, i.e, technique (c) in the list above, the control-flow model discovered in the first time window $W_1$, is also obtained via Inductive Miner~\cite{Leemans13}, while the control-flow model was subsequently adapted, as per Step 1 discussed in Section~\ref{sec:approach}. For the other perspectives, we leverage on  Hoeffding Adaptive Trees, for which we set a maximum depth of $5$, as done for techniques (a) and (b). Moreover, we experimented with various \textit{grace periods} ($100$, $500$, $1000$, $5000$, $10000$, $50000$), which determine the number of instances observed before considering a split at a node, and using in the simulation model the one with best accuracy. Particularly, lower grace periods facilitate rapid adaption to sudden concept drifts, while higher grace periods  reduce the risk of overfitting and promote more stable decisions. 
%\color{blue}
Details of the experiments using fixed grace periods are in Appendix~\ref{sec:ap:metrics_gp}.

\subsection{Results}
\label{sec:results}
To assess the accuracy of our proposed technique, we ran simulations for each time window and computed the distances between the event logs obtained via the different simulation techniques, and the original ones. Distances were computed using the metrics proposed in~\cite{CHAPELACAMPA2025102447}. Specifically, CFLD and 3GD assess control-flow related distances, AED and RED measure number of events distributions over time, in absolute or relative terms, respectively. CED and CWD evaluate the simulator's ability to accurately replicate events within the circadian dimension (day and hour of the week). CAR quantifies the accuracy of case arrivals, and CTD measures the difference in cycle time distributions, implicitly capturing the accuracy of execution waiting times~\cite{CHAPELACAMPA2025102447}.

\begin{table}[t!]
    \centering
    \caption{Average results of simulations between time windows per each technique. Numbers in parentheses indicate standard deviations. 
    The online technique results are highlighted with a green background. 
    % For each measure, the average results for each method are reported. The best results are highlighted in bold.
    }
    \resizebox{0.83\columnwidth}{!}{%
    \begin{tabular}{|c|c||c|c|>{\columncolor{green!20}}c|}
    \hline
        \textbf{Measure} & \textbf{Event Log} & \textbf{Single Batch} & \textbf{Last Batch} & \textbf{Online} \\
    \hline\hline
    \multirow{5}{*}{CFLD} & BPIC17W & $\mathbf{0.16}\ (0.05)$ & $0.17\ (0.04)$ & $0.18\ (\mathbf{0.02})$ \\
    & BPIC12W & $0.25\ (0.08)$ & $0.42\ (0.07)$ & $\mathbf{0.17}\ (\mathbf{0.05})$ \\
    & P2P     & $\mathbf{0.2}\ (\mathbf{0.03})$ & $0.39\ (0.18)$ & $\mathbf{0.2}\ (\mathbf{0.03})$ \\
    & CVS & $\mathbf{0.02}\ (\mathbf{0.0})$ & $0.07\ (0.08)$ & $\mathbf{0.02}\ (\mathbf{0.0})$ \\
    \cline{2-5}
    & Avg. & $ 0.16\ (0.04) $ & $ 0.26\ (0.09) $ & $\mathbf{0.14}\ (\mathbf{0.03})$\\
    \hline\hline
    \multirow{5}{*}{3GD} & BPIC17W & $\mathbf{0.18}\ (0.07)$ & $0.19\ (0.05)$ & $0.19\ (\mathbf{0.04})$ \\
    & BPIC12W & $0.26\ (0.11)$ & $0.54\ (\mathbf{0.06})$ & $\mathbf{0.25}\ (0.09)$ \\
    & P2P     & $\mathbf{0.22}\ (\mathbf{0.02})$ & $0.4\ (0.19)$ & $\mathbf{0.22}\ (\mathbf{0.02})$ \\
    & CVS & $\mathbf{0.03}\ (\mathbf{0.0})$ & $0.1\ (0.1)$ & $\mathbf{0.03}\ (\mathbf{0.0})$ \\
    \cline{2-5}
    & Avg. & $\mathbf{0.17}\ (0.05)$ & $0.31\ (0.1)$ & $\mathbf{0.17}\ (\mathbf{0.04})$\\
    \hline\hline
    \multirow{5}{*}{AED} & BPIC17W & $1205.86\ (643.01)$ & $1260.28\ (\mathbf{579.05})$ & $\mathbf{1100.72}\ (680.09)$ \\
    & BPIC12W & $736.2\ (530.88)$ & $717.86\ (548.07)$ & $\mathbf{687.44}\ (\mathbf{518.25})$ \\
    & P2P     & $1475\ (622.17)$ & $\mathbf{869.08}\ (\mathbf{504.36})$ & $1132.41\ (626.99)$ \\
    & CVS & $60.63\ (24.5)$ & $5711.09\ (10952.39)$ & $\mathbf{38.54}\ (\mathbf{16.89})$ \\
    \cline{2-5}
    & Avg. & $869.42\ (\mathbf{455.14})$ & $2139.58\ (3145.97)$ & $\mathbf{739.78}\ (460.56)$\\
    \hline\hline
    \multirow{5}{*}{RED} & BPIC17W & $41.52\ (21.68)$ & $83.74\ (22.57)$ & $\mathbf{24.01}\ (\mathbf{11.11})$ \\
    & BPIC12W & $72.96\ (28.24)$ & $156.32\ (41.07)$ & $\mathbf{44.61}\ (\mathbf{22.36})$ \\
    & P2P     & $616.65\ (270.55)$ & $676.77\ (351.08)$ & $\mathbf{535.81}\ (\mathbf{252.49})$ \\
    & CVS & $51.95\ (13.74)$ & $47.4\ (10.0)$ & $\mathbf{20.19}\ (\mathbf{3.0})$ \\
    \cline{2-5}
    & Avg. & $195.77\ (83.55)$ & $241.06\ (106.18)$ & $\mathbf{156.16}\ (\mathbf{72.24})$\\
    \hline\hline
    \multirow{5}{*}{CED} & BPIC17W & $1.38\ (0.9)$ & $1.42\ (1.22)$ & $\mathbf{0.82}\ (\mathbf{0.61})$ \\
    & BPIC12W & $2.89\ (1.22)$ & $2.45\ (1.55)$ & $\mathbf{2.45}\ (\mathbf{1.5})$ \\
    & P2P     & $\mathbf{0.88}\ (0.22)$ & $2.34\ (0.86)$ & $1.06\ (\mathbf{0.15})$ \\
    & CVS & $0.27\ (0.15)$ & $1.06\ (0.6)$ & $\mathbf{0.13}\ (\mathbf{0.03})$ \\
    \cline{2-5}
    & Avg. & $1.36\ (0.62)$ & $1.82\ (1.06)$ & $\mathbf{1.12}\ (\mathbf{0.57})$\\
    \hline\hline
    \multirow{5}{*}{CWD} & BPIC17W     & $1.25\ (0.87)$ & $1.33\ (1.18)$ & $\mathbf{0.69}\ (\mathbf{0.39})$ \\
    & BPIC12W & $2.77\ (\mathbf{1.16})$ & $\mathbf{2.27}\ (1.53)$ & $2.37\ (1.51)$ \\
    & P2P     & $\mathbf{0.85}\ (\mathbf{0.13})$ & $2.13\ (0.89)$ & $0.96\ (0.14)$ \\
    & CVS & $0.22\ (0.06)$ & $0.71\ (0.54)$ & $\mathbf{0.19}\ (\mathbf{0.05})$ \\
    \cline{2-5}
    & Avg. & $1.27\ (0.56)$ & $1.61\ (1.04)$ & $\mathbf{1.05}\ (\mathbf{0.52})$\\
    \hline\hline
    \multirow{5}{*}{CAR} & BPIC17W & $1262.21\ (678.0)$ & $1272.08\ (\mathbf{574.62})$ & $\mathbf{1241.19}\ (761.5)$ \\
    & BPIC12W & $791.45\ (628.59)$ & $\mathbf{718.37}\ (\mathbf{524.64})$ & $778.33\ (590.46)$ \\
    & P2P     & $1025.01\ (\mathbf{431.99})$ & $\mathbf{617.11}\ (478.91)$ & $799.21\ (452.99)$ \\
    & CVS & $25.28\ (18.66)$ & $5759.53\ (10963.3)$ & $\mathbf{18.51}\ (\mathbf{13.66})$ \\
    \cline{2-5}
    & Avg. & $775.99\ (\mathbf{439.31})$ & $2091.77\ (3135.37)$ & $\mathbf{709.31}\ (454.65)$\\
    \hline\hline
    \multirow{5}{*}{CTD} & BPIC17W     & $63.6\ (41.46)$ & $99.07\ (34.12)$ & $\mathbf{36.82}\ (\mathbf{17.78})$ \\
    & BPIC12W  & $94.26\ (41.86)$ & $194.03\ (37.21)$ & $\mathbf{60.39}\ (\mathbf{23.75})$\\
    & P2P     & $729.88\ (357.63)$ & $900.86\ (504.84)$ & $\mathbf{647.69}\ (\mathbf{294.34})$\\
    & CVS & $95.75\ (27.28)$ & $85.95\ (23.21)$ & $\mathbf{41.07}\ (\mathbf{3.66})$\\
    \cline{2-5}
    & Avg. & $245.87\ (117.06)$ & $319.98\ (149.85)$ & $\mathbf{196.49}\ (\mathbf{84.88})$\\
    \hline
    \end{tabular}
    }
    \label{tab:results}
\end{table}

For each use case, we conducted five simulations and computed the average distances between the obtained event logs for each time window. The final results are presented in \autoref{tab:results}, where the column with the green background highlights our technique. The reported values represent the average distance across all time windows, with standard deviations provided in parentheses. Since they are distances, lower average values indicate better performances. Additionally, lower standard deviation values suggest greater stability and consistency  in accuracy across the temporal dimension.

The results demonstrate that in general our technique outperforms the baselines. Indeed, we achieve equal or superior average results across all metrics, highlighting its effectiveness. Notably, the last batch technique consistently yields poor average results, emphasizing the importance of not just ignoring older portions of event-data sets.
Looking at the control-flow measures (CFLD and 3GD), we can notice that they are very close to those obtained using the single batch technique. This suggests that no significant process changes were detected in these control-flow metrics. 
Very good results are achieved with the Cycle Time Distribution (CTD) and the Relative Event Distribution (RED) distances, consistently outperforming the other techniques. These results indicate that our technique effectively adapts to temporal perspective changes.

Analyzing individual event logs and metrics, our technique achieves the best results in $24$ out of $32$ cases (i.e. $75\%$ of the cases), compared to $8$ cases ($25\%$) for the single batch technique and only $4$ cases ($12.5\%$) for the last batch technique. Note that there are $4$ cases where our technique and the single batch technique yield identical average results.
Standard deviation analysis further supports the robustness of our technique. In $24$ out of $32$ cases, our technique demonstrates greater stability and consistency over time compared to the other two techniques.

\begin{figure}[t!]
    \centering
    \includegraphics[width=\linewidth]{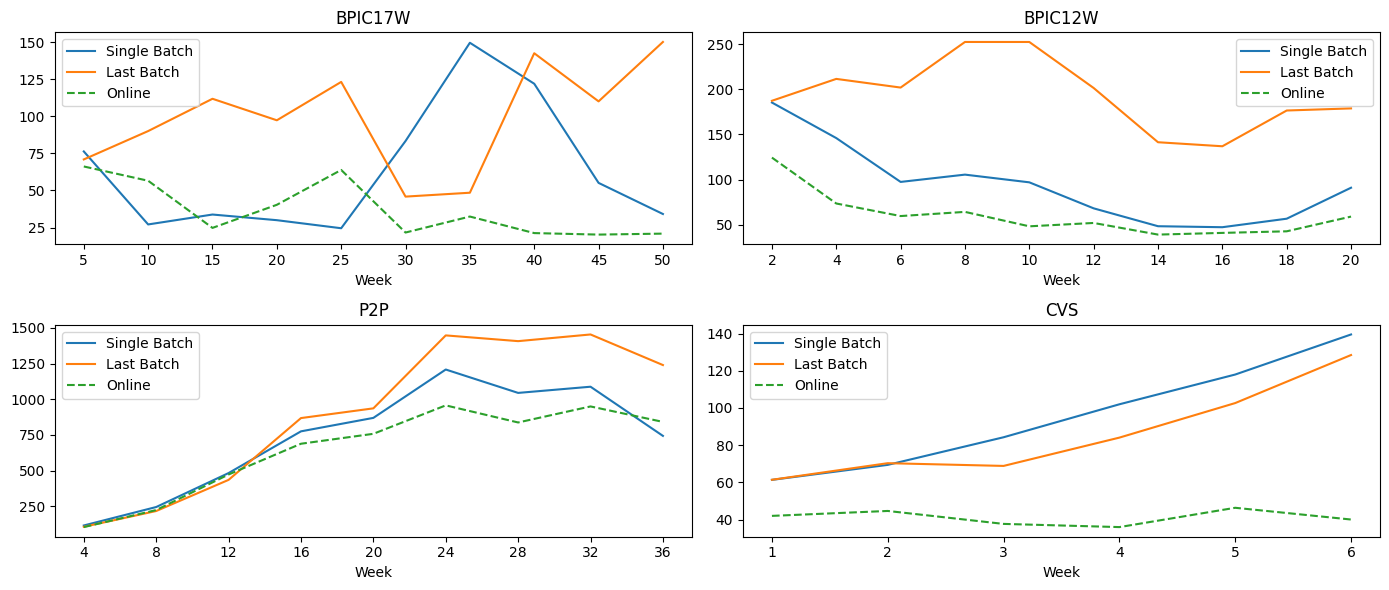}
    \caption{Cycle Time Distribution (CTD) distance over time for each use case.}
    \label{fig:ctd}
\end{figure}

We also visually explored the results through time windows.~\autoref{fig:ctd} illustrates the Cycle Time Distribution (CTD) distance over time for each use case. Our technique consistently outperforms the others in BPIC12W and CVS. In some cases, such as BPIC17W and CVS, the last batch technique yields better results than the single batch technique. Overall, our technique demonstrates greater stability, producing more consistent results over time.
% Plots for each case study are available in the Appendix of the paper extended version available in~\cite{extended_vers}.
Plots for each metric are available in Appendix~\ref{sec:ap:metrics}.

\begin{figure}[t!]
\centering
\begin{subfigure}[t!]{.48\textwidth}
  \centering
  \includegraphics[width=\linewidth]{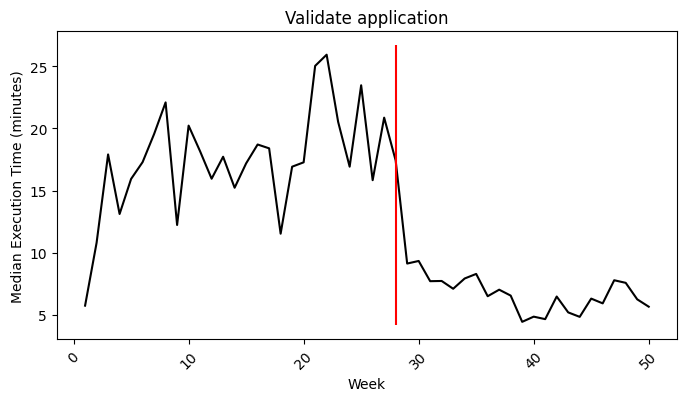}
  \caption{Median duration over weeks (in minutes) of the \emph{Validate Application} activity.\newline}
  \label{fig:validate_concept_drift}
\end{subfigure}
\hfill
\begin{subfigure}[t!]{.48\textwidth}
  \centering
  \includegraphics[width=\linewidth]{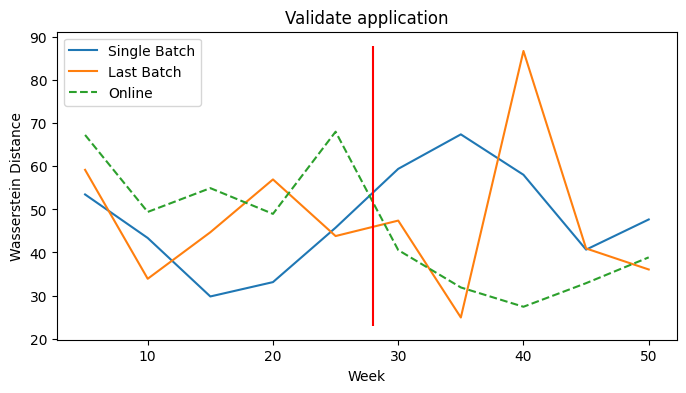}
  \caption{Wasserstein distances between the actual distribution of durations and those obtained via simulations over weeks.}
  \label{fig:validate_wd}
\end{subfigure}
\vspace{5pt}
\caption{\textit{Validate application} execution time results. The vertical red lines indicate when the concept drift has been detected in~\cite{Adams2021}.}
\label{fig:res_concept_drift}
\end{figure}

The BPIC17W use case was one for which our technique has most remarkably outperformed the baseline, especially in the comparison metrics related to the time and resource perspectives. Previous work has indeed highlighted a presence of significant drift related to the increase of the resource workload~\cite{Adams2021}. For this case study, we conducted a more thorough assessment related to activity \emph{Validate Application}: the average duration of the activity shows a reduction of $38\%$ at week 28 (see \autoref{fig:validate_concept_drift}).
In particular, we computed the Wasserstein distance between the simulated and real execution time distributions for \textit{Validate application}. \autoref{fig:validate_wd} shows these distances over weeks, where one could clearly see that our technique certainly leads to lower Wasserstein distances after the concept drift, except for two weeks, if compared with the baseline techniques. 
This also implicitly contributes to better CTD results after week 28 (see~\autoref{fig:ctd}).

\section{Conclusion}
\label{sec:conclusion}
Traditional Business Process Simulation discovery techniques rely on analyzing finite sets of historical traces. These methods assume the process does not change over time, treating both past and recent event behaviors equally. However, real-world business process can evolve over time due to internal policies changes aimed at process improvement or external influencing factors. In this paper, we proposed a simulation discovery technique able to adapt to evolving processes.

State-of-the-art research has demonstrated that hybrid process simulation models can produce more accurate simulations~\cite{CAMARGO2023102248}. To maintain high accuracy, we combined Incrementally Process Discovery~\cite{danielThesis} with Online Machine Learning~\cite{hoeffding} techniques. As new process instances are executed, the simulation model is updated. The control-flow perspective of the simulation model is updated using the technique proposed by Schuster et al.~\cite{IPD_trace_fragments}, while the other perspectives rely on predictive models that are based on Hoeffding Adaptive Trees~\cite{hoeffding}, which can adaptively learn from data streams that evolve over time.

The conducted experiments (cf. Section~\ref{sec:exp}) reveal the potential effectiveness of our technique. 
The evaluation uses four distinct processes and their associated event logs. 
% To analyze temporal evolution, the logs were segmented into ten sequential windows, denoted $W_1, \ldots, W_{10}$, where $t_i$ represents the latest timestamp of $W_i$. Three process simulation models were generated for each timestamp $t_i$: (a) using all traces contained from batches $W_1$ to $W_i$, (b) using only traces from $W_i$, and (c) employing an incremental technique across $W_1$ to $W_i$. The synthetic event logs produced by these three models at each $t_i$ were then compared against the actual traces starting in batch $W_{i+1}$ using established performance metrics.
The results show that our technique can potentially lead to more accurate results across various perspectives, and it is more stable over time.

%\color{blue}
%\subsection{Limitations and Future works}
We acknowledge that the work by Schuster et al.~\cite{IPD_trace_fragments} does not explicitly remove unobserved behaviors from the process model, and this is an open challenge in the fields of Incremental Process Discovery and Repair. However, this does not pose major challenges in our studies, since we rely on predictive models to estimate branching probabilities at decision points, rare or unobserved paths are naturally assigned near-zero probabilities. However, we acknowledge that the readability of the models would be improved, if those “unused” parts were not in the model. We plan to work on this as an avenue of future work, especially in a event-log streaming settings. 
%\color{black}Finally, we also aim to explore different types of process changes, such as seasonal variations, through time series analysis, and integrate specific models for these patterns within BPS models.

% \begin{credits}
% \subsubsection{\ackname} A bold run-in heading in small font size at the end of the paper is
% used for general acknowledgments, for example: This study was funded

% \subsubsection{\discintname}
% It is now necessary to declare any competing interests or to specifically
% state that the authors have no competing interests. Please place the
% statement with a bold run-in heading in small font size beneath the
% (optional) acknowledgments\footnote{If EquinOCS, our proceedings submission
% system, is used, then the disclaimer can be provided directly in the system.},
% for example: The authors have no competing interests to declare that are
% relevant to the content of this article. Or: Author A has received research
% grants from Company W. Author B has received a speaker honorarium from
% Company X and owns stock in Company Y. Author C is a member of committee Z.
% \end{credits}

\begin{small}
\paragraph{Acknowledgments.} F.\ Vinci is financially supported by MUR (PNRR) and University of Padua. M.\ de Leoni is supported by the European Union – Next Generation EU under the National Recovery and Resilience Plan (NRRP), Mission 4 Component 2 Investment 1.1 - Call PRIN 2022 PNRR No. 1409 of September 14, 2022 of Italian Ministry of University and Research; Project P20222XM58 (subject area: SH)\textit{ Emergency medicine 4.0: an integrated data-driven approach to improve emergency department performances}.
\end{small}

\bibliographystyle{splncs04}
\bibliography{biblio}

\begin{thebibliography}{10}
\providecommand{\url}[1]{\texttt{#1}}
\providecommand{\urlprefix}{URL }
\providecommand{\doi}[1]{https://doi.org/#1}

\bibitem{vanderAalst2016}
van~der Aalst, W.: Process Mining: Data Science in Action. Springer Berlin Heidelberg (2016). \doi{10.1007/978-3-662-49851-4\_1}

\bibitem{Adams2021}
Adams, J.N., van Zelst, S.J., Quack, L., Hausmann, K., van~der Aalst, W.M.P., Rose, T.: A framework for explainable concept drift detection in process mining. In: Business Process Management. pp. 400--416. Springer International Publishing, Cham (2021). \doi{10.1007/978-3-030-85469-0_25}

\bibitem{hoeffding}
Bifet, A., Gavald{\`a}, R.: Adaptive learning from evolving data streams. In: Advances in Intelligent Data Analysis VIII. pp. 249--260. Springer Berlin Heidelberg, Berlin, Heidelberg (2009). \doi{10.1007/978-3-642-03915-7_22}

\bibitem{Bose}
Bose, R.P.J.C., van~der Aalst, W.M.P., Žliobaitė, I., Pechenizkiy, M.: Dealing with concept drifts in process mining. IEEE Transactions on Neural Networks and Learning Systems  \textbf{25}(1),  154--171 (2014). \doi{10.1109/TNNLS.2013.2278313}

\bibitem{Burattin2022}
Burattin, A.: Streaming process mining. In: Process Mining Handbook, pp. 349--372. Springer International Publishing, Cham (2022). \doi{10.1007/978-3-031-08848-3_11}

\bibitem{Burattin2014}
Burattin, A., Sperduti, A., van~der Aalst, W.M.P.: Control-flow discovery from event streams. In: 2014 IEEE Congress on Evolutionary Computation (CEC). pp. 2420--2427 (2014). \doi{10.1109/CEC.2014.6900341}

\bibitem{CAMARGO2023102248}
Camargo, M., Báron, D., Dumas, M., González-Rojas, O.: Learning business process simulation models: A hybrid process mining and deep learning approach. Information Systems  \textbf{117},  102248 (2023). \doi{10.1016/j.is.2023.102248}

\bibitem{CAMARGO2020113284}
Camargo, M., Dumas, M., González-Rojas, O.: Automated discovery of business process simulation models from event logs. Decision Support Systems  \textbf{134},  113284 (2020). \doi{10.1016/j.dss.2020.113284}

\bibitem{Carmona2012}
Carmona, J., Gavald{\`a}, R.: Online techniques for dealing with concept drift in process mining. In: Advances in Intelligent Data Analysis XI. pp. 90--102. Springer Berlin Heidelberg, Berlin, Heidelberg (2012). \doi{10.1007/978-3-642-34156-4_10}

\bibitem{CHAPELACAMPA2025102447}
Chapela-Campa, D., Benchekroun, I., Baron, O., Dumas, M., Krass, D., Senderovich, A.: A framework for measuring the quality of business process simulation models. Information Systems  \textbf{127},  102447 (2025). \doi{10.1016/j.is.2024.102447}

\bibitem{Klijn2024}
Klijn, E.L., Mannhardt, F., Fahland, D.: Multi-perspective concept drift detection: Including the actor perspective. In: Advanced Information Systems Engineering. pp. 141--157. Springer Nature Switzerland, Cham (2024). \doi{10.1007/978-3-031-61057-8_9}

\bibitem{Kraus2025}
Kraus, A., van~der Aa, H.: Machine learning-based detection of concept drift in business processes. Process Science  \textbf{2} (2025). \doi{10.1007/s44311-025-00012-w}

\bibitem{Leemans13}
Leemans, S.J.J., Fahland, D., van~der Aalst, W.M.P.: Discovering block-structured process models from event logs - a constructive approach. In: Application and Theory of Petri Nets and Concurrency. pp. 311--329. Springer Berlin Heidelberg, Berlin, Heidelberg (2013). \doi{10.1007/978-3-642-38697-8_17}

\bibitem{deLeoniBPM2023}
de~Leoni, M., Vinci, F., Leemans, S.J.J., Mannhardt, F.: Investigating the influence of data-aware process states on activity probabilities in simulation models: Does accuracy improve? In: Business Process Management. pp. 129--145. Springer Nature Switzerland (2023). \doi{10.1007/978-3-031-41620-0_8}

\bibitem{Lu2021}
Lu, Y., Chen, Q., Poon, S.: A robust and accurate approach to detect process drifts from event streams. In: Business Process Management. pp. 383--399. Springer International Publishing, Cham (2021). \doi{10.1007/978-3-030-85469-0_24}

\bibitem{MENEGHELLO2025102472}
Meneghello, F., {D}i Francescomarino, C., Ghidini, C., Ronzani, M.: Runtime integration of machine learning and simulation for business processes: Time and decision mining predictions. Information Systems  \textbf{128},  102472 (2025). \doi{10.1016/j.is.2024.102472}

\bibitem{river}
Montiel, J., Halford, M., Mastelini, S.M., Bolmier, G., Sourty, R., Vaysse, R., Zouitine, A., Gomes, H.M., Read, J., Abdessalem, T., Bifet, A.: River: machine learning for streaming data in python. Journal of Machine Learning Research  \textbf{22} (2021)

\bibitem{Navarin2020}
Navarin, N., Cambiaso, M., Burattin, A., Maggi, F.M., Oneto, L., Sperduti, A.: Towards online discovery of data-aware declarative process models from event streams. In: 2020 International Joint Conference on Neural Networks (IJCNN). pp.~1--8 (2020). \doi{10.1109/IJCNN48605.2020.9207500}

\bibitem{Rizzi_2022}
Rizzi, W., Di Francescomarino, C., Ghidini, C., Maggi, F.M.: How do i update my model? on the resilience of predictive process monitoring models to change. Knowledge and Information Systems  \textbf{64}(5),  1385–1416 (Mar 2022). \doi{10.1007/s10115-022-01666-9}

\bibitem{ROZINAT2009305}
Rozinat, A., Mans, R., Song, M., {van der Aalst}, W.: Discovering simulation models. Information Systems  \textbf{34}(3),  305--327 (2009). \doi{10.1016/j.is.2008.09.002}

\bibitem{Sato2021}
Sato, D.M.V., De~Freitas, S.C., Barddal, J.P., Scalabrin, E.E.: A survey on concept drift in process mining. ACM Comput. Surv.  \textbf{54}(9) (2021). \doi{10.1145/3472752}

\bibitem{Scheibel2023}
Scheibel, B., Rinderle-Ma, S.: An end-to-end approach for online decision mining and decision drift analysis in process-aware information systems. In: Intelligent Information Systems. pp. 17--25. Springer International Publishing, Cham (2023). \doi{10.1007/978-3-031-34674-3_3}

\bibitem{danielThesis}
Schuster, D.: {I}ncremental process discovery. Dissertation, RWTH Aachen University, Aachen (2024). \doi{10.18154/RWTH-2024-06483}

\bibitem{IPD_trace_fragments}
Schuster, D., F{\"o}cking, N., van Zelst, S.J., van~der Aalst, W.M.P.: Incremental discovery of process models using trace fragments. In: Business Process Management. pp. 55--73. Springer Nature Switzerland, Cham (2023). \doi{10.1007/978-3-031-41620-0_4}

\bibitem{cortado}
Schuster, D., {van Zelst}, S.J., {van der Aalst}, W.M.: Cortado: A dedicated process mining tool for interactive process discovery. SoftwareX  \textbf{22},  101373 (2023). \doi{10.1016/j.softx.2023.101373}

\bibitem{basThesis}
{van Zelst}, S.: Process mining with streaming data. Phd thesis (2019)

\bibitem{Bas2018}
van Zelst, S.J., Dongen, B.F., van~der Aalst, W.M.: Event stream-based process discovery using abstract representations. Knowledge and Information System  \textbf{54}(2),  407–435 (2018). \doi{10.1007/s10115-017-1060-2}

\end{thebibliography}

\newpage
\appendix
\section{Distances Over Time}\label{sec:ap:metrics}
\begin{figure}
    \centering
    \includegraphics[width=\linewidth]{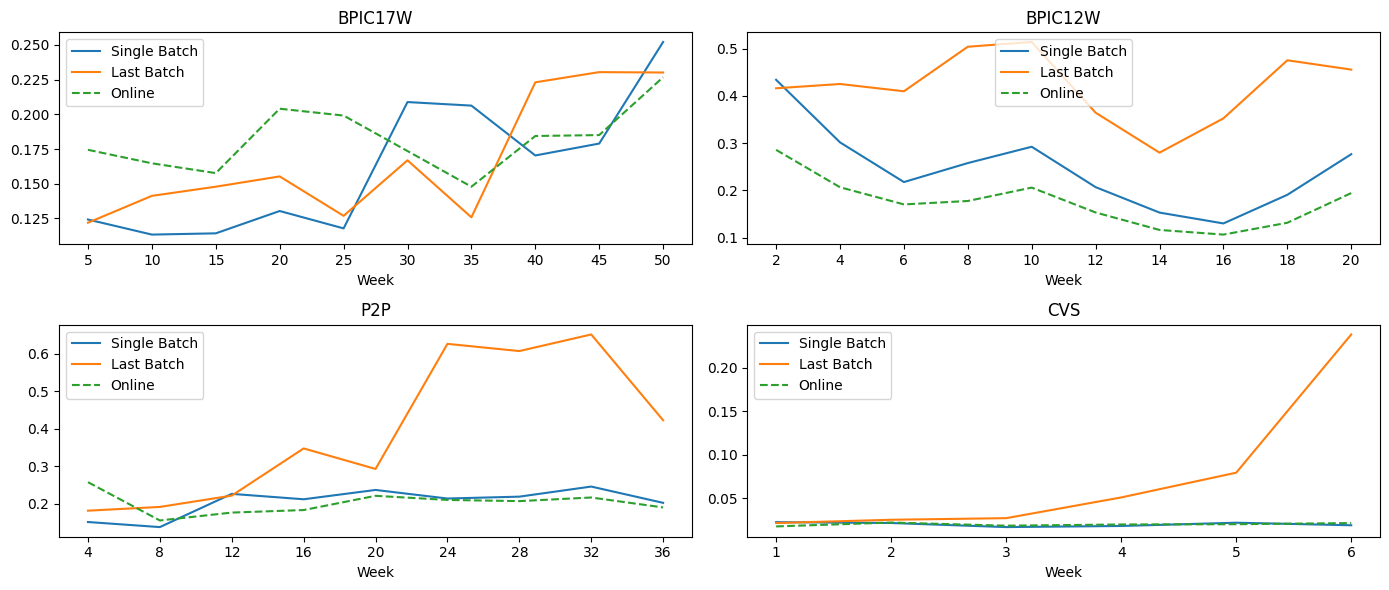}
    \caption{Control-Flow Log Distance (CFLD) over time for each use case.}
    \label{fig:cfld}
\end{figure}

\begin{figure}
    \centering
    \includegraphics[width=\linewidth]{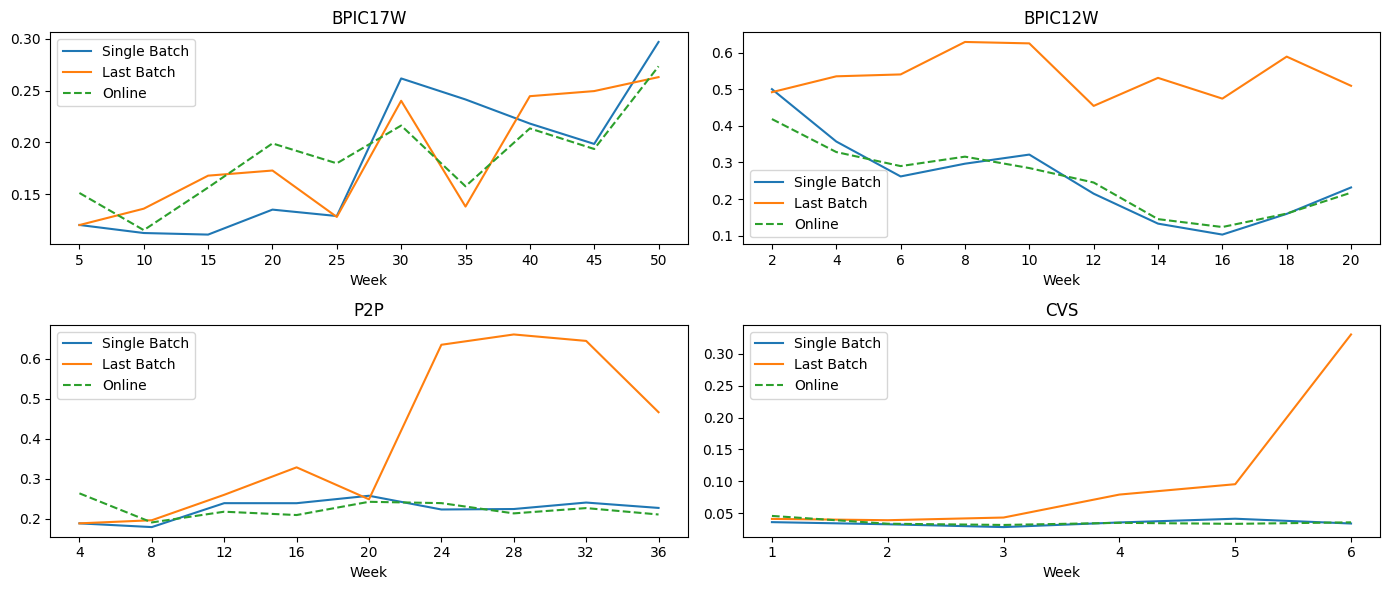}
    \caption{3-Gram Distance (3DG) over time for each use case.}
    \label{fig:ngd}
\end{figure}

\begin{figure}
    \centering
    \includegraphics[width=\linewidth]{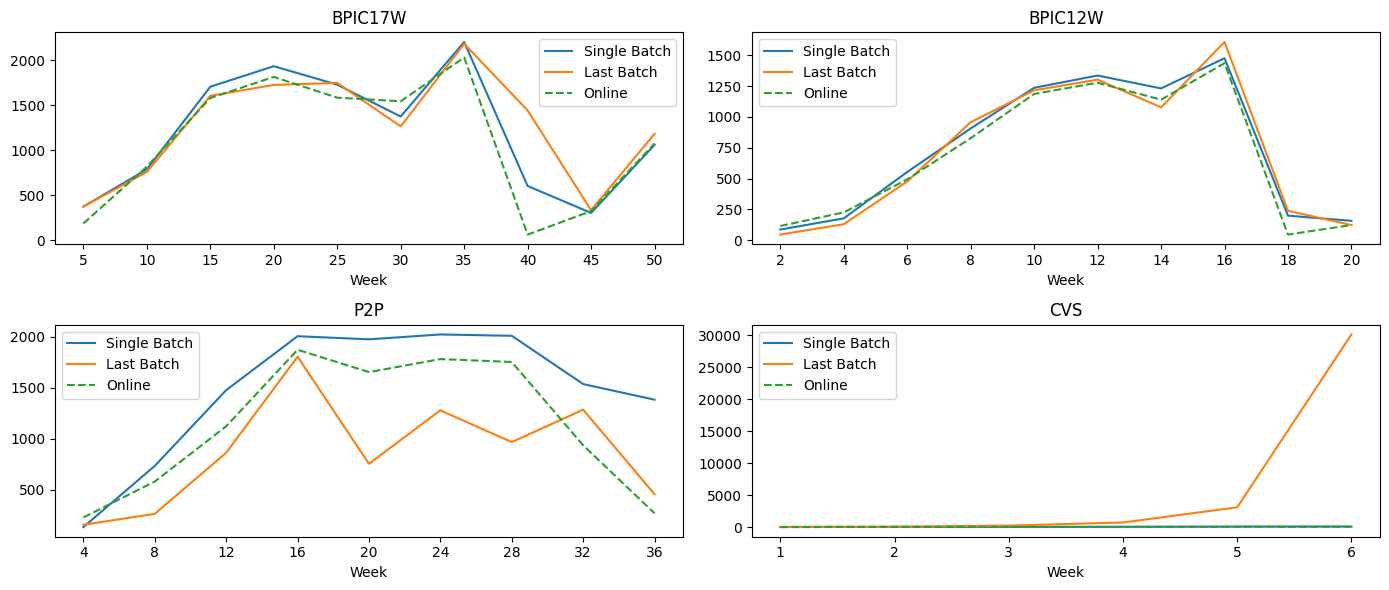}
    \caption{Absolute Event Distribution (AED) distance over time for each use case.}
    \label{fig:aed}
\end{figure}

\begin{figure}
    \centering
    \includegraphics[width=\linewidth]{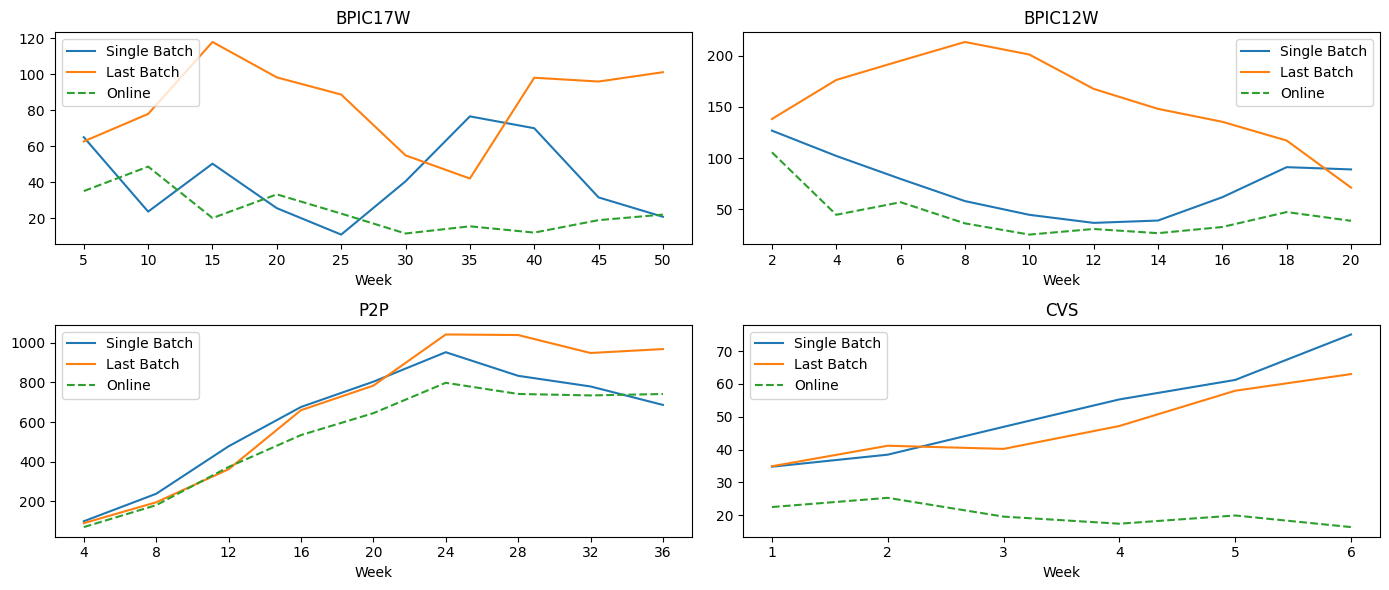}
    \caption{Relative Event Distribution (RED) distance over time for each use case.}
    \label{fig:red}
\end{figure}

\begin{figure}
    \centering
    \includegraphics[width=\linewidth]{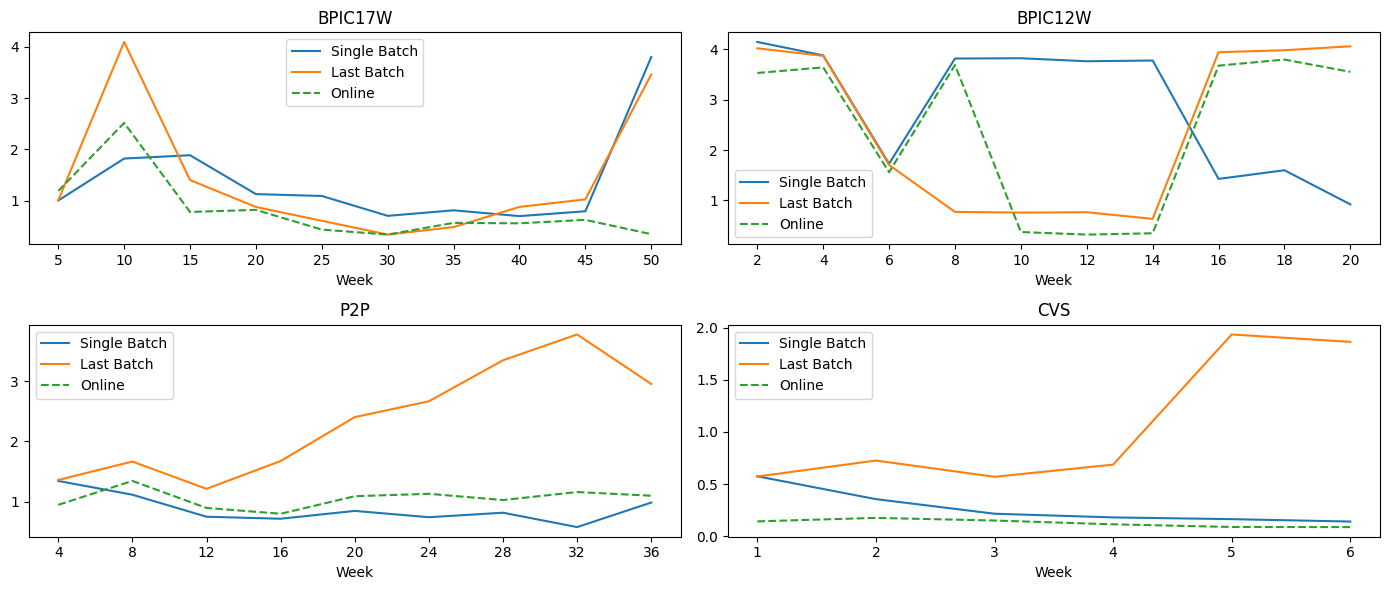}
    \caption{Circadian Event Distribution (CED) distance over time for each use case.}
    \label{fig:ced}
\end{figure}

\begin{figure}
    \centering
    \includegraphics[width=\linewidth]{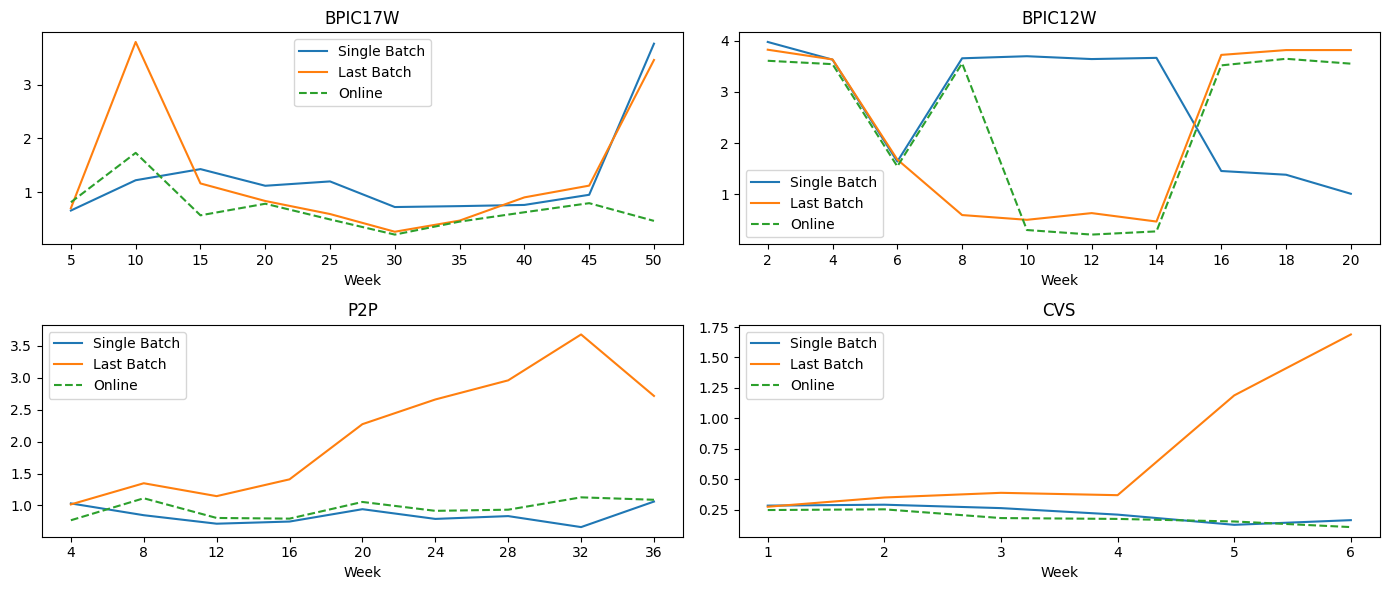}
    \caption{Circadian Workload Distribution (CWD) distance over time for each use case.}
    \label{fig:cwd}
\end{figure}

\begin{figure}
    \centering
    \includegraphics[width=\linewidth]{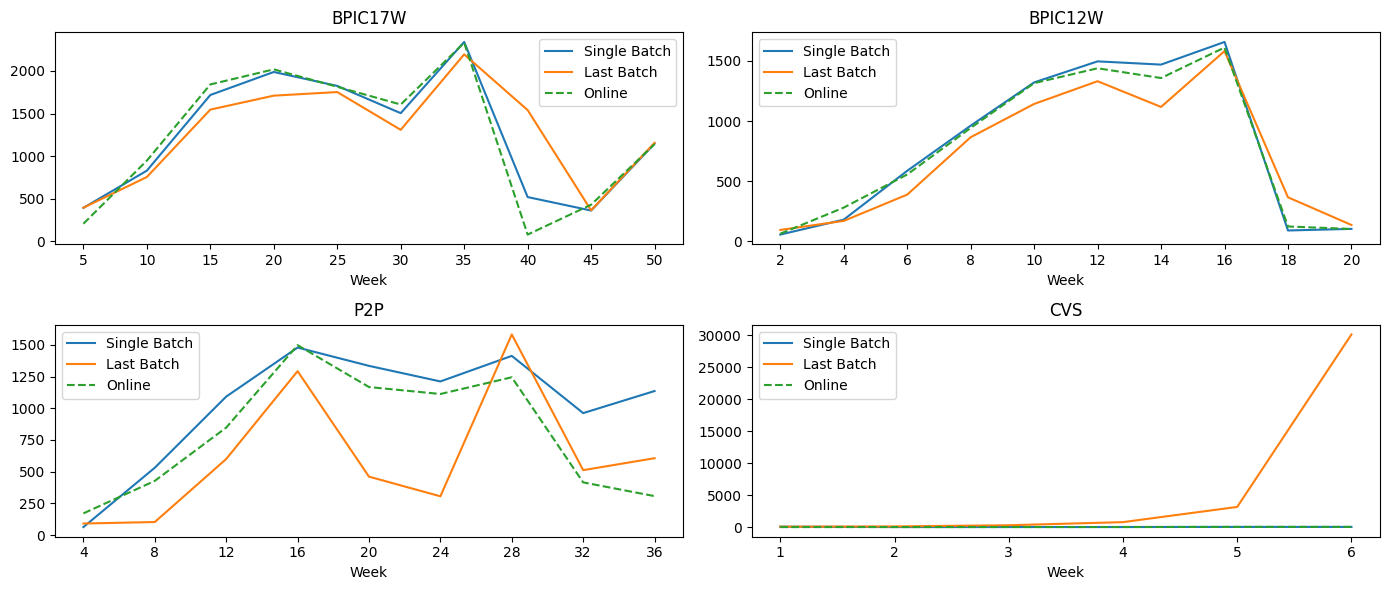}
    \caption{Case Arrival Rate (CAR) distance over time for each use case.}
    \label{fig:car}
\end{figure}

\begin{figure}
    \centering
    \includegraphics[width=\linewidth]{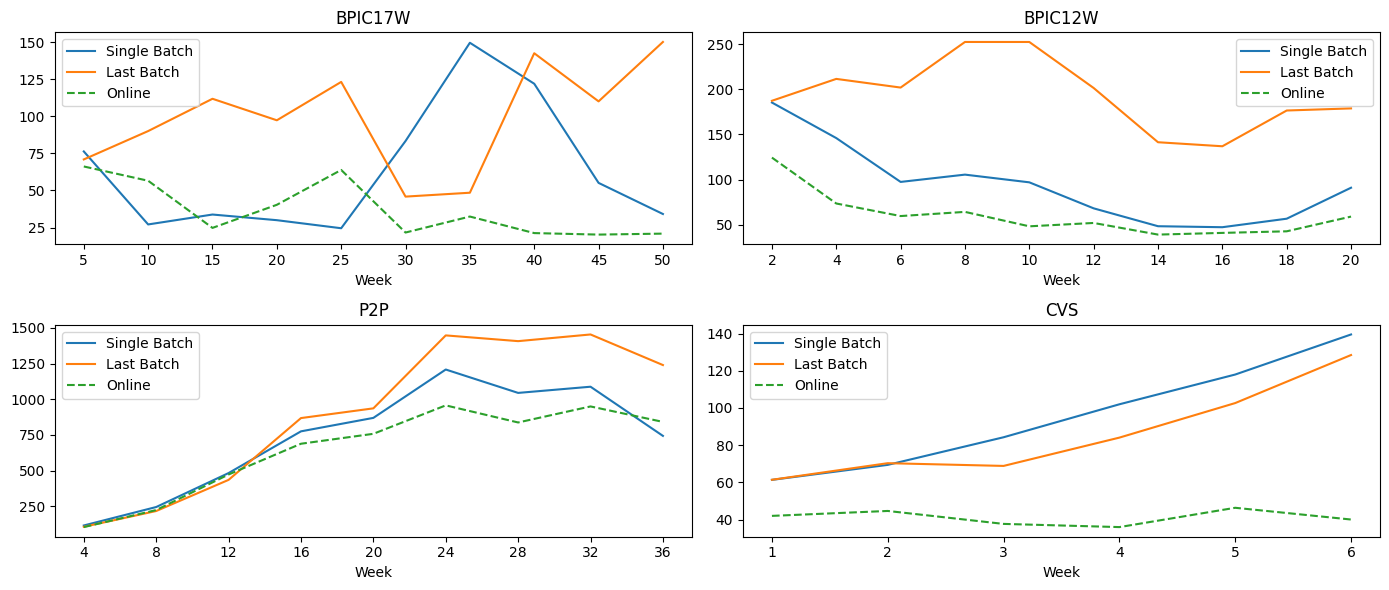}
    \caption{Cycle Time Distribution (CTD) distance over time for each use case.}
    \label{fig:ctd_app}
\end{figure}

This section presents detailed visualizations for each evaluation metric in all use cases. Figures~\ref{fig:cfld}--\ref{fig:ctd_app} offer a comprehensive overview of how metrics evolve over time for each technique (cf. Section~\ref{sec:results}). In particular, they illustrate how the online method generally produces more accurate and stable performance across metrics.

\newpage
\section{Distances Over Time per Grace Period}\label{sec:ap:metrics_gp}
\begin{figure}
    \centering
    \includegraphics[width=\linewidth]{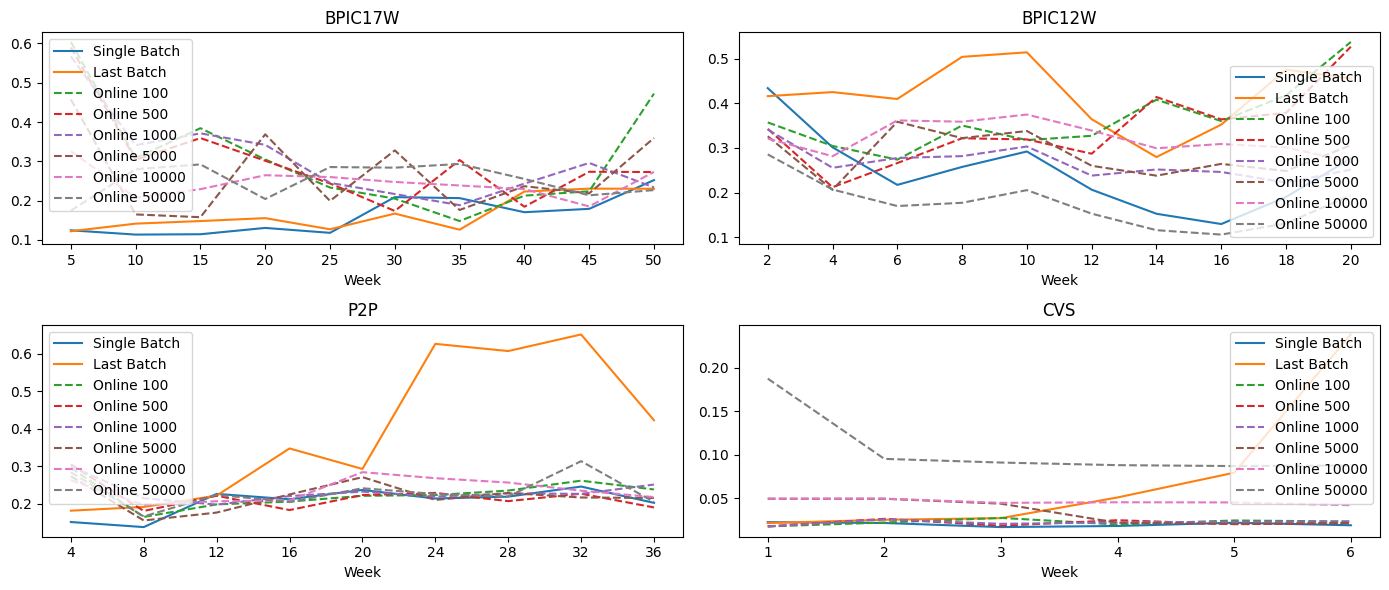}
    \caption{Control-Flow Log Distance (CFLD) over time for each use case per grace period.}
    \label{fig:cfld_gp}
\end{figure}

\begin{figure}
    \centering
    \includegraphics[width=\linewidth]{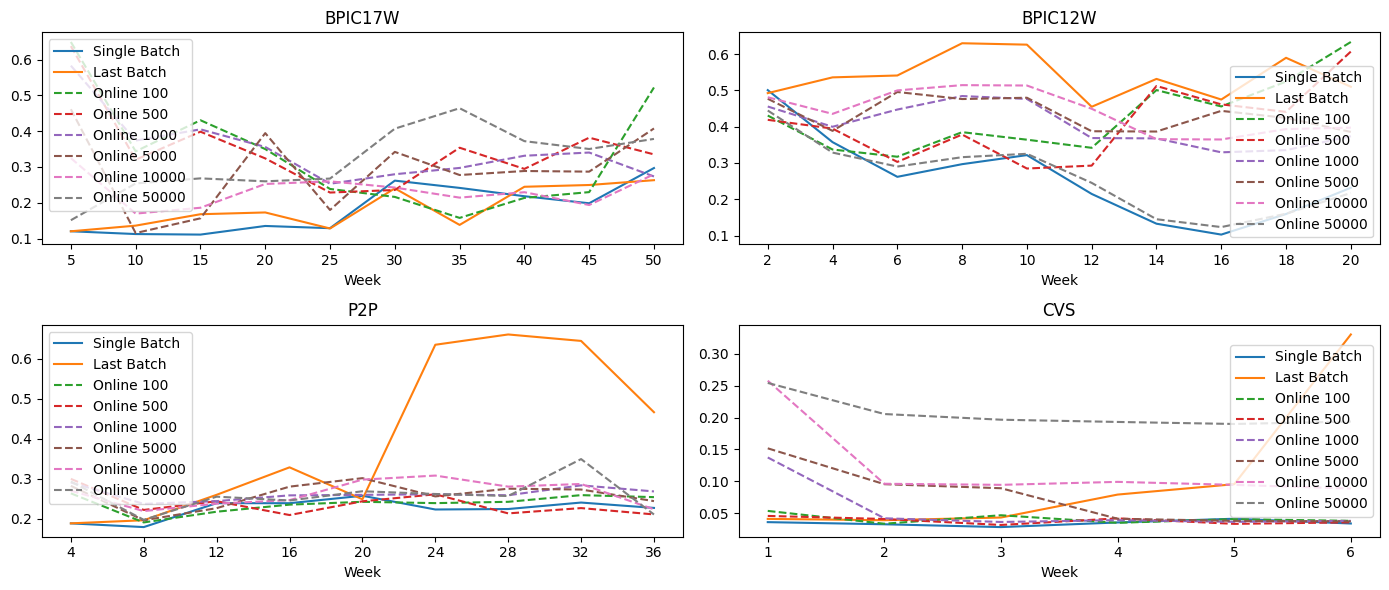}
    \caption{3-Gram Distance (3DG) over time for each use case per grace period.}
    \label{fig:ngd_gp}
\end{figure}

\begin{figure}
    \centering
    \includegraphics[width=\linewidth]{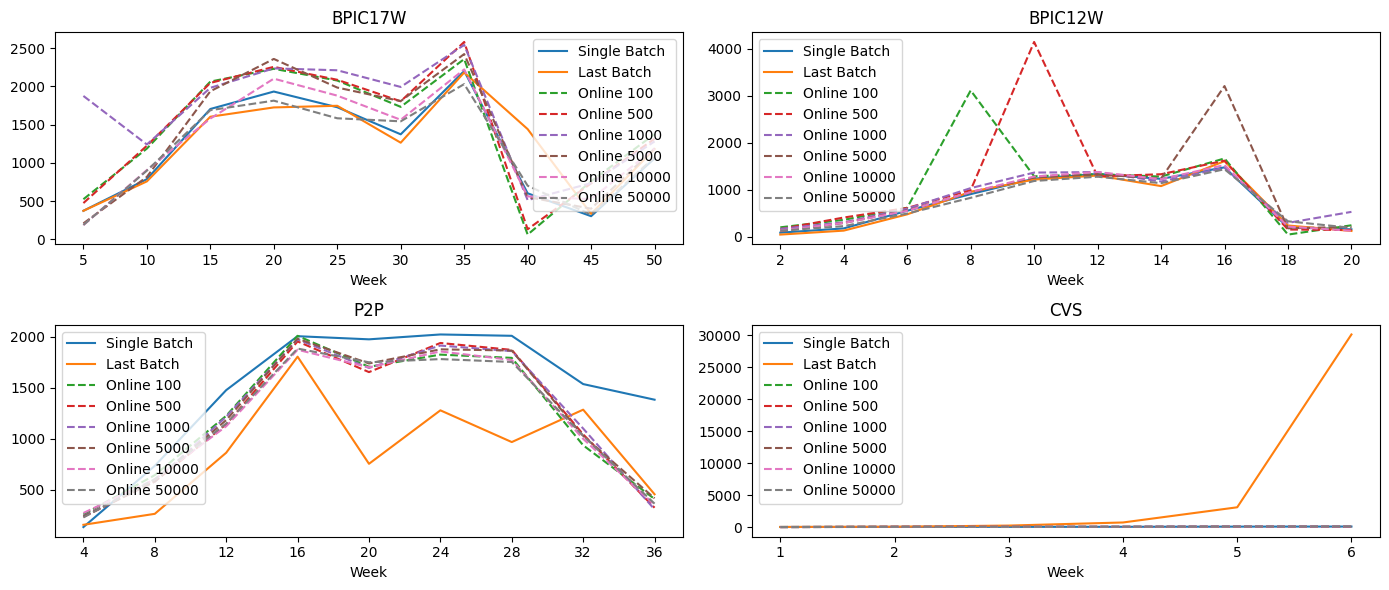}
    \caption{Absolute Event Distribution (AED) distance over time for each use case per grace period.}
    \label{fig:aed_gp}
\end{figure}

\begin{figure}
    \centering
    \includegraphics[width=\linewidth]{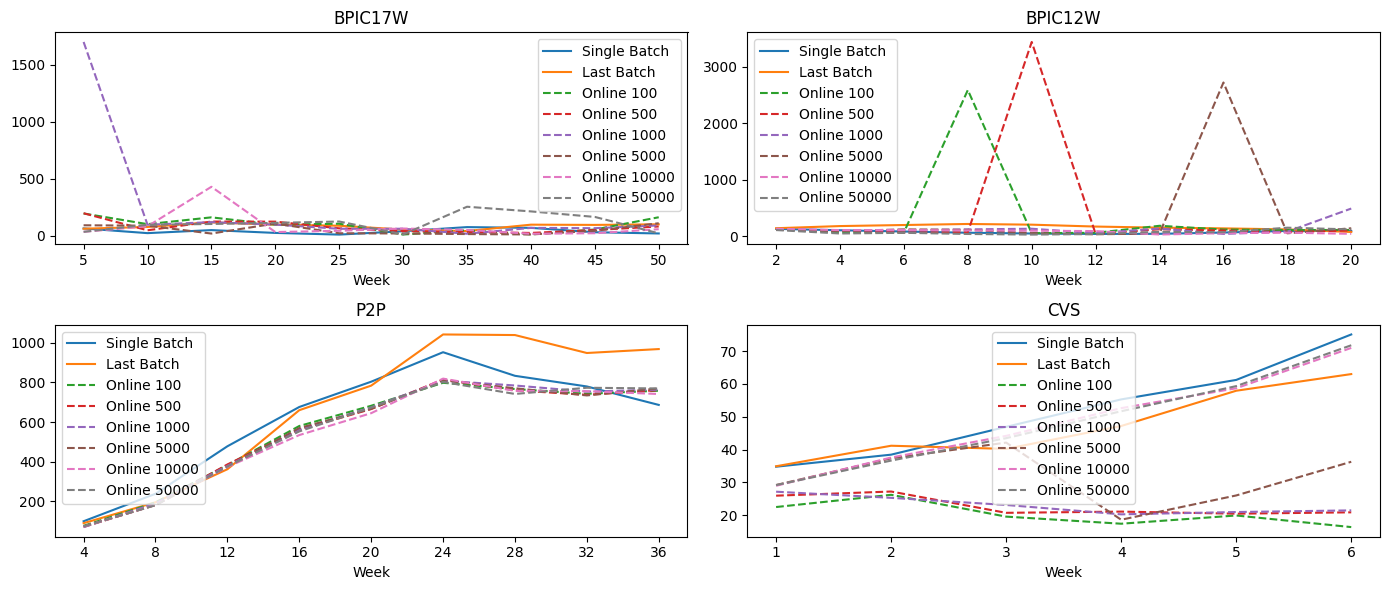}
    \caption{Relative Event Distribution (RED) distance over time for each use case per grace period.}
    \label{fig:red_gp}
\end{figure}

\begin{figure}
    \centering
    \includegraphics[width=\linewidth]{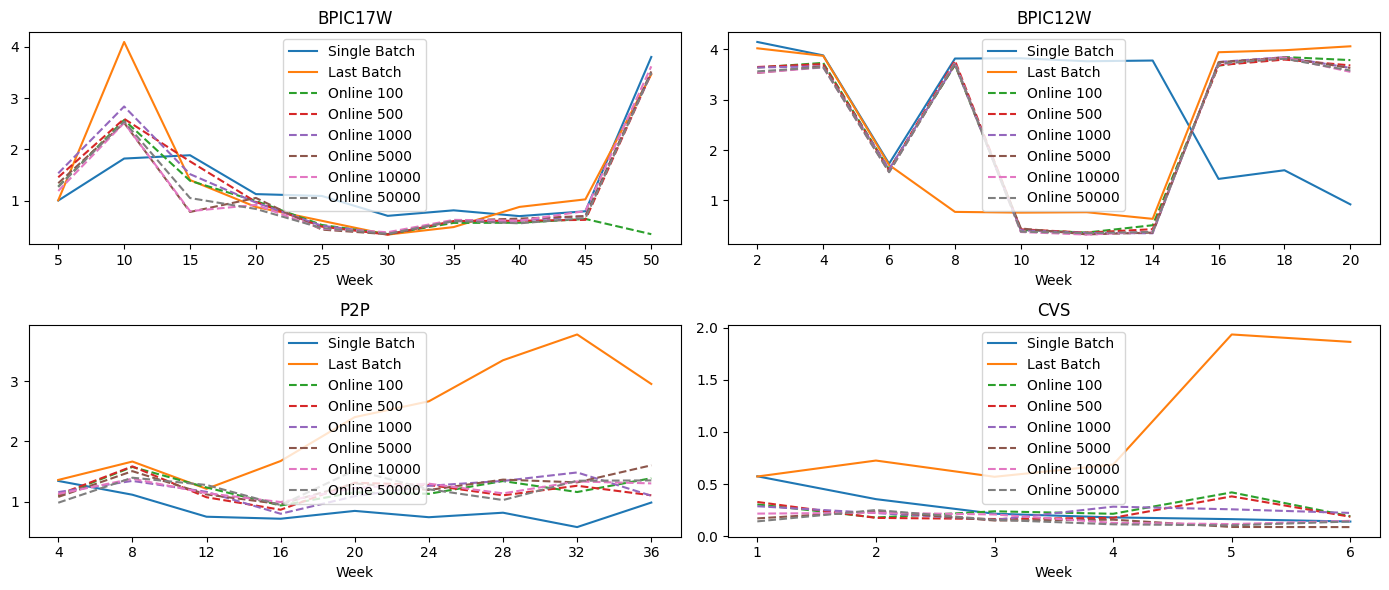}
    \caption{Circadian Event Distribution (CED) distance over time for each use case per grace period.}
    \label{fig:ced_gp}
\end{figure}

\begin{figure}
    \centering
    \includegraphics[width=\linewidth]{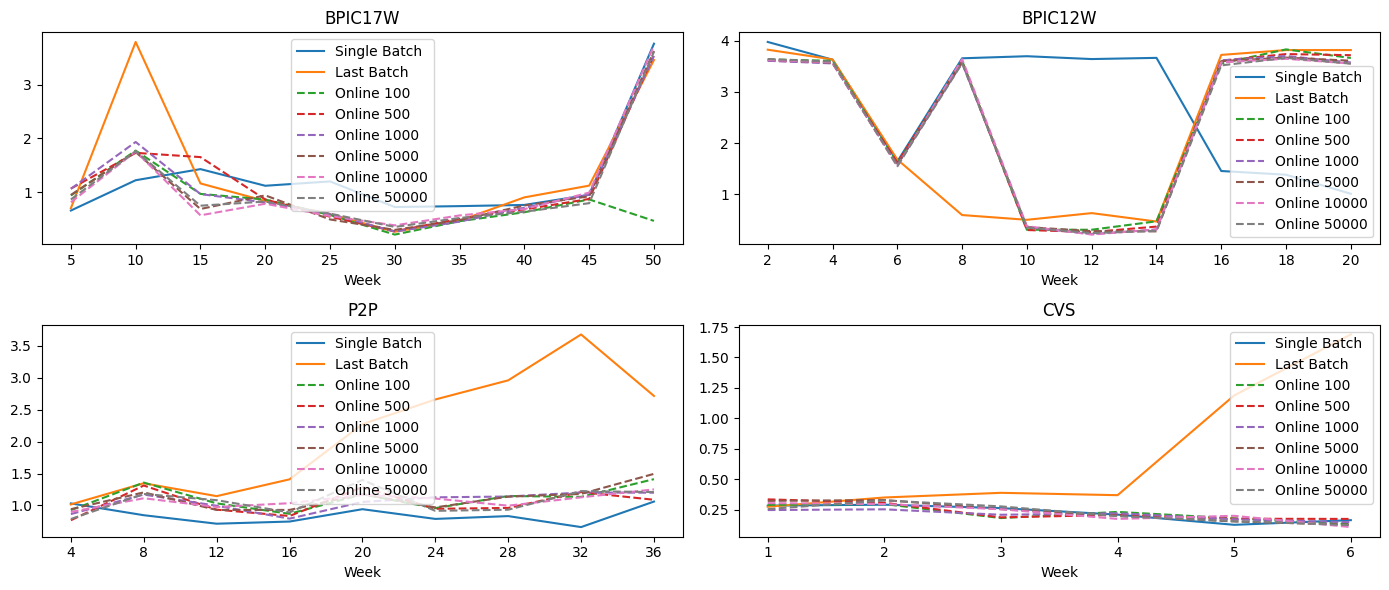}
    \caption{Circadian Workload Distribution (CWD) distance over time for each use case per grace period.}
    \label{fig:cwd_gp}
\end{figure}

\begin{figure}
    \centering
    \includegraphics[width=\linewidth]{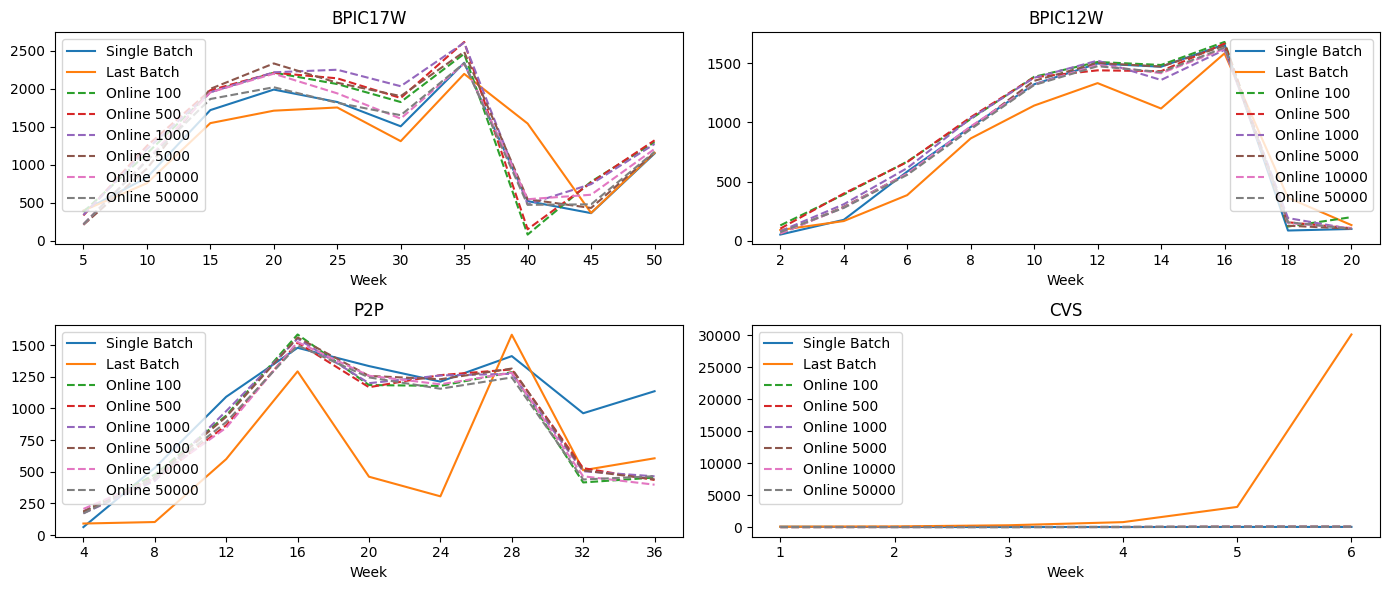}
    \caption{Case Arrival Rate (CAR) distance over time for each use case per grace period.}
    \label{fig:car_gp}
\end{figure}

\begin{figure}
    \centering
    \includegraphics[width=\linewidth]{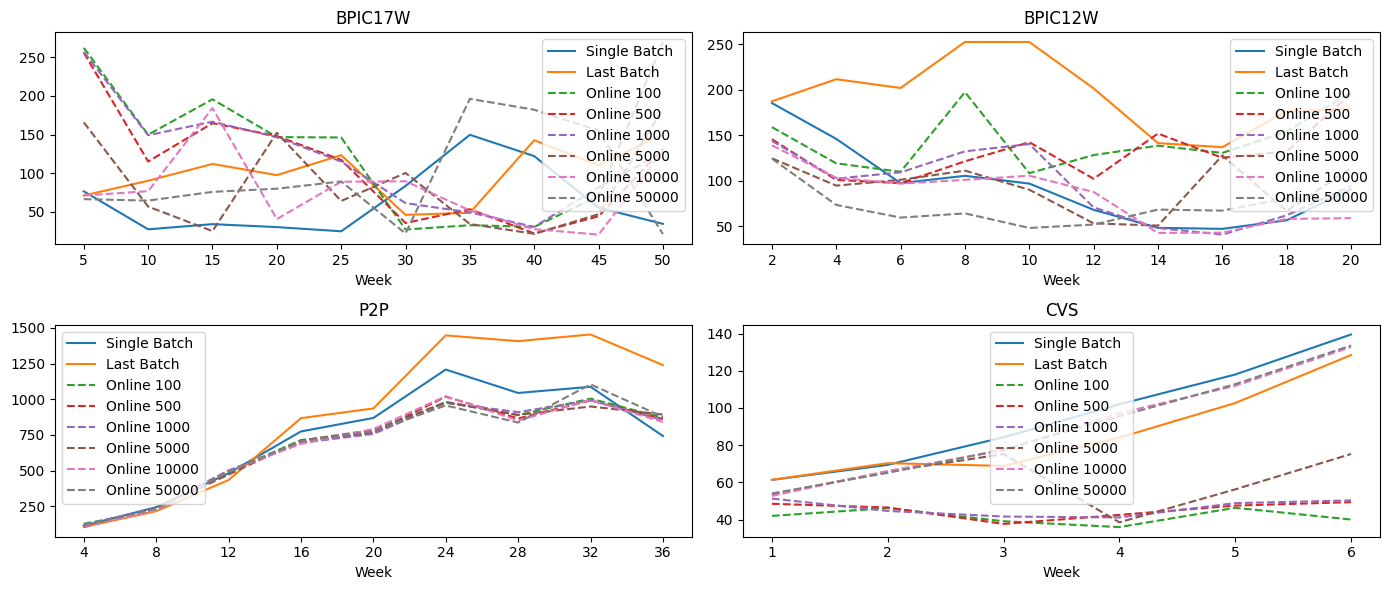}
    \caption{Cycle Time Distribution (CTD) distance over time for each use case per grace period.}
    \label{fig:ctd_app_gp}
\end{figure}

This Appendix provides a detailed analysis of the experiments conducted using fixed grace periods of $100$, $500$, $1000$, $5000$, $10000$, and $50000$. These experiments aimed to evaluate the impact of the grace period parameter on model performance across different scenarios. Figures\ref{fig:cfld_gp}--\ref{fig:ctd_app_gp} illustrate how varying the grace period affects the balance between adaptability and stability. Particularly, lower grace periods facilitate rapid adaption to sudden concept drifts, while higher grace periods reduce the risk of overfitting and promote more stable decisions. The results guided the selection of the optimal grace period used in the simulation model.

\end{document}